\documentclass[12pt, a4paper, axodraw]{article}
\pdfoutput=1
\usepackage[english]{babel}
\usepackage{graphicx}
\usepackage[onehalfspacing]{setspace}
\usepackage{geometry}
\usepackage{indentfirst}
\usepackage{anyfontsize}
\graphicspath{ {figures/} }
\usepackage{array}
\usepackage{float}
\usepackage{caption}
\usepackage{graphicx}
\usepackage{amsmath}
\usepackage[RPvoltages]{circuitikz}
\usepackage{subcaption}
\usepackage[utf8]{inputenc}
\usepackage{hyperref}
\hypersetup{
    colorlinks=true,
    linkcolor=blue,
    filecolor=magenta,      
    urlcolor=cyan,
    citecolor=cyan
}
\usepackage{systeme}
\usepackage{physics}
\usepackage{bm}
\usepackage{wrapfig}
\usepackage{mathtools}
\usepackage{amssymb , amsthm}
\usepackage{bbold}
\usepackage{amsmath}
\usepackage{ragged2e}
\usepackage{paralist}
\usepackage{blindtext}
\usepackage{hyperref}
\usepackage[T1]{fontenc}
\usepackage{mathptmx}
\hypersetup{urlcolor=blue}
\usepackage{graphicx}
\DeclareMathAlphabet{\mathcal}{OMS}{cmsy}{m}{n}

\begin{document}
\begin{center}
    \vspace{4cm}
    
    \textbf{\Huge Study of CP Violation in charm decays}
    \vspace{2cm}
    
    \textbf{Author:} Felipe Cesário Laterza Lopes$^1$
    
    \vspace{0.5cm}
    
    \textbf{Advisor:} Prof. Dr. Patrícia Camargo Magalhães$^1$

    \vspace{0.5cm}

    \textit{\ $^1$Institute of Physics "Gleb Wataghin", UNICAMP, Campinas, São Paulo, Brasil}
    \vspace{1cm}

    \begin{abstract}
        The violation of the Charge-Parity (CP) symmetry is a phenomenon described by the Standard Model (SM), however, its predictions for this violation are not enough to explain the matter-antimatter asymmetry in the visible universe. Experimentally, CP violation in hadronic decays had already been observed in various circumstances before 2019, when the LHCb experiment at CERN first published results reporting CPV in D-meson decays\textsuperscript{\cite{observation_cpv}} ($D^0 \rightarrow K^- K^+$ and $D^0 \rightarrow \pi^- \pi^+$). In this work, a study of CP violation is carried out analysing the phenomenology of charm decays. The study of such topic is guided by the Cabibbo-Kobayashi-Maskawa (CKM) matrix and the process of flavour changing decays caused by the weak interaction (focusing on charm decays). The symmetries attributed to physical conservation laws were recognized, allowing for the possibilities of CP violation predicted by the SM to be examined more closely. Such theoretical studies are then compared to the results obtained by the LHCb Collaboration. Furthermore, a comparison between CPV in charm decays and in neutrino oscillations is made, analysing the CP violating phase in both the CKM and the Pontecorvo–Maki–Nakagawa–Sakata (PMNS) matrices.
    \end{abstract}
\end{center}

\vfill
\newpage

{\hypersetup{linkcolor=black} \tableofcontents}

\vfill
\newpage

\newcommand{\bigi}{\textsuperscript{\cite{bigi_sanda_2009}}}
\newcommand{\griffiths}{\textsuperscript{\cite{Griffiths_2008}}}
\newcommand{\thomson}{\textsuperscript{\cite{Thomson_2013}}}
\newcommand{\article}{\textsuperscript{\cite{observation_cpv}}}
\newcommand{\pdg}{\textsuperscript{\cite{pdgbooklet}}}
\newcommand{\giunti}{\textsuperscript{\cite{giunti}}}

\section{Introduction}

The Standard Model (SM) is the theory of particle physics that describes the dynamics of the matter that composes the known universe, and it is the most accurate model to our current understanding in regards to the description of elementary particles and the way they interact\thomson. In order to accomplish such task, it describes the fundamental fermions that define the most fundamental particles, as well as the fundamental bosons, which are responsible for the interactions between the particles.

When particles interact, some symmetries are often observed. Since 1964, several experimental observations have been made of the charge conjugation and parity (CP) symmetries being violated in different physical processes, more notably in kaon systems in 1964 and B-meson systems in 2001. The breaking of this symmetry is one of the conditions stipulated by Sakharov in 1967 for baryogenesis, that would explain the matter-antimatter asymmetry in the observable universe. 

One such occasion when CP is violated is during meson decays, through quark flavour changing processes that depends on the complex phase of the Cabibbo-Kobayashi-Maskawa (CKM) matrix. This phenomenon had been observed experimentally in kaon and B-meson system. However, in 2019, the LHCb collaboration announced the first direct observations of CP violation in D-meson systems. In other words, that was the first experimental evidence for CP violation in the decay of the charm quark.

Hence, we will seek to introduce the violation of the charge-parity symmetry by studying its sources within the standard model. With such purpose, we will discuss its origin in the weak interaction and analyse it through a description of the decay of the $D^0$ meson. In order to contextualize this description, a comparison between the theoretical models and the known experimental results will also be carried out.

By the end of this work it will be possible to understand the different mechanisms of CP violation, as well as its correlation to the complex phase in the CKM matrix and how we can generalize this mechanism to other structures such as neutrino oscillations described by the Pontecorvo–Maki–Nakagawa–Sakata (PMNS) matrix.

\section{Standard Model}

\begin{wrapfigure}{R}{0.4\linewidth}
    \centering
    \captionsetup{justification=centering}
    \includegraphics[width=0.39\textwidth]{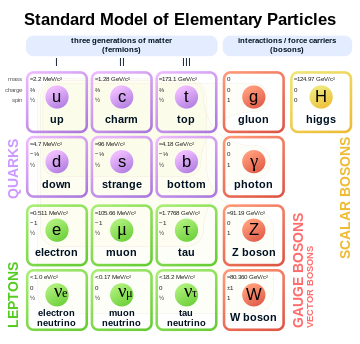}
    \caption{Particles described by the Standard Model of particle physics}
    \label{fig:SM}
\end{wrapfigure}

In order to understand how a symmetry can be violated in particle interactions, we must first understand what those particles are and how they interact.
The Standard Model defines two types of fundamental particles: fermions (with half-integer spin) and bosons (with integer spin). Fermions are the ones that compose what we usually call "matter", whereas bosons are responsible for the interactions between fermions. The reason for such is that in the SM, forces are described by the exchange of bosons between fermions\thomson. The fermions are separated in two categories: quarks and leptons (with both being separated in three generations), meanwhile bosons are separated into gauge bosons and scalar bosons (Figure \ref{fig:SM}).

Among the particles described in Figure \ref{fig:SM}, few comprise the matter we see around us: the atoms that form such matter are bound states of electrons ($e^-$), protons and neutrons. These last two are in turn composed by a combination of up and down quarks\thomson. Due to small lifetimes or small probability of interacting (as is the case of neutrinos), fermions of the second and third generations are often harder to observe. Nonetheless, them and their interactions are essential to understand what is observed at higher energies.

The four known fundamental forces are strong, electromagnetic, weak and gravitational\griffiths. However, when discussing elementary particle dynamics, we only take in consideration the first three due to their relative strength: the effect of gravity in any elementary particle can be safely disregarded for it is many orders of magnitude smaller than any other. 

Therefore, by considering the other three forces, we can very accurately describe the processes undergone by elementary particles. The three relevant forces are described by a Quantum Field Theory (QFT) associated to the exchange of a gauge boson. The electromagnetic interaction can be described by Quantum Electrodynamics (QED), which describes the exchange of photons. The strong force is described by Quantum Chromodynamics (QCD), with its gauge boson being the gluon. Finally, the weak interaction is mediated by the exchange of two charged-current bosons ($W^+$ and $W^-$) and one neutral-current boson ($Z^0$)\thomson. 

Independently of which interactions they go through, the dynamics of fermions is described by the Dirac equation. Similarly to the way the Schrödinger Equation describes the dynamics of particles in non-relativistic quantum mechanics, the Dirac Equation is responsible for the description of particles with spin 1/2 in relativistic quantum mechanics. One of the consequences of the Dirac Equation is that its solution requires the existence of what we now call an "antiparticle": an antiparticle state has the same mass but opposite charge to its equivalent particle state\thomson. The notation of antiparticles for leptons consists on inverting its charge (the positron $e^+$ is the antiparticle of the electron $e^-$) and for quarks it consists of a bar over the particle symbol (the anti-charm quark is $\bar{q}$)\thomson. 

Unlike leptons, quarks and antiquarks can interact through the three forces. They carry electric charge (and thus take part in the electromagnetic interaction), but also carry a "color" quantum number related to the strong interaction, described by QCD. This is motivated by the fact that gluons are responsible for the "exchange" of color between particles. However, we only observe "colorless" particles as free particles in nature, and since quarks carry one of three possible color charges (and antiquarks carry an anti-color), they combine into colorless particles by grouping a quark-antiquark pair or by grouping three quarks(antiquarks), each with one of the colors(anti-colors). These bound states called hadrons\thomson. 

Due to the three color charges, these hadrons can be categorized into baryons and mesons. Typical baryons consist of three colored quarks and have their corresponding anti-baryon, whereas mesons are composed by one quark and one antiquark. The particles involved in the process we wish to study are all mesons, and in Figure \ref{tab:part} we list these particles and the quarks which constitute them. 

\begin{wrapfigure}{r}{0.3\textwidth}
    \centering
    \begin{tabular}{|c|c|}
        \hline
        Meson & Structure \\
        \hline
        $D^0$ & $c\bar{u}$ \\
        $\bar{D}^0$ & $u\bar{c}$ \\
        $K^+$ & $u\bar{s}$ \\
        $K^-$ & $s\bar{u}$ \\
        $\pi^+$ & $u\bar{d}$ \\
        $\pi^-$ & $d\bar{u}$ \\
        \hline
    \end{tabular}
    \caption{Particle listing}
    \label{tab:part}
\end{wrapfigure}

As described by the SM through the Quantum Field Theories, particles interact via coupling to force-carrying bosons\thomson only if the particle itself carries the charge of the interaction. Such charge for the charged-current weak interaction is known as weak isospin, which is carried by all fermions, thus allowing all fermions to interact via the weak force. Furthermore, since the weak charged-current bosons ($W^+$ and $W^-$) carry electric charge of $+e$ and $-e$ respectively, they only couple pairs of fundamental particles that differ by one unit of electric charge. Weak neutral currents are well known, but are not involved in the $D^0$ meson interactions we wish to discuss, hence we shall attain ourselves to a description of weak charged-current interactions.

\subsection{Weak Interaction}
\label{sec:weak}

The $W$ bosons can couple to leptons only within the same family (equation \ref{eq:leptonfamilies})\thomson, conserving electron, muon and tau numbers,

\begin{equation}
    \begin{pmatrix}
    \nu_e  \\
    e^-  
    \end{pmatrix},
    \begin{pmatrix}
    \nu_{\mu}  \\
    \mu^-  
    \end{pmatrix},
    \begin{pmatrix}
    \nu_{\tau}  \\
    \tau^-  
    \end{pmatrix}.
    \label{eq:leptonfamilies}
\end{equation}

However, these bosons can also couple to quarks, causing interactions in which quark flavors do not need to be conserved, as has been observed experimentally. The different quark generations are\griffiths 

\begin{equation}
    \begin{pmatrix}
    u  \\
    d  
    \end{pmatrix},
    \begin{pmatrix}
    c  \\
    s  
    \end{pmatrix},
    \begin{pmatrix}
    t  \\
    b  
    \end{pmatrix}.
    \label{eq:quarkfamilies}
\end{equation}

If couplings were restricted to same generation quarks, particles as the $K^-$ and $B$ mesons would be absolutely stable, which has been observed to not be the case. Hence, cross-generational couplings are also allowed, still restricted to quarks differing by one unit charge\thomson:

\begin{equation}
    \begin{pmatrix}
    u  \\
    s  
    \end{pmatrix},
    \begin{pmatrix}
    c  \\
    d  
    \end{pmatrix},
    \begin{pmatrix}
    t  \\
    d  
    \end{pmatrix},
    \begin{pmatrix}
    u  \\
    b  
    \end{pmatrix},
    \begin{pmatrix}
    c  \\
    d  
    \end{pmatrix},
    \begin{pmatrix}
    t  \\
    s  
    \end{pmatrix}.
    \label{eq:quarkcgcoupling}
\end{equation}

Since $W^{\pm}$ can couple to both quarks and leptons, we can categorize decays by which fermions are involved in the interaction. If the $W^{\pm}$ is only coupled to leptons, we call it a leptonic process, if it couples to quarks at one vertex and leptons at the other, it is a semileptonic process. And, if the coupling is exclusive to quarks, it is a purely hadronic process\griffiths.

The decays of the $D^0$ meson we wish to discuss are purely hadronic processes which involve cross-generational couplings and can be represented by their Feynman diagrams. 

\begin{figure}[H]
    \vspace{-0.5cm}
    \centering
    \begin{subfigure}[b]{0.49\textwidth}
        \includegraphics[width=0.9\linewidth]{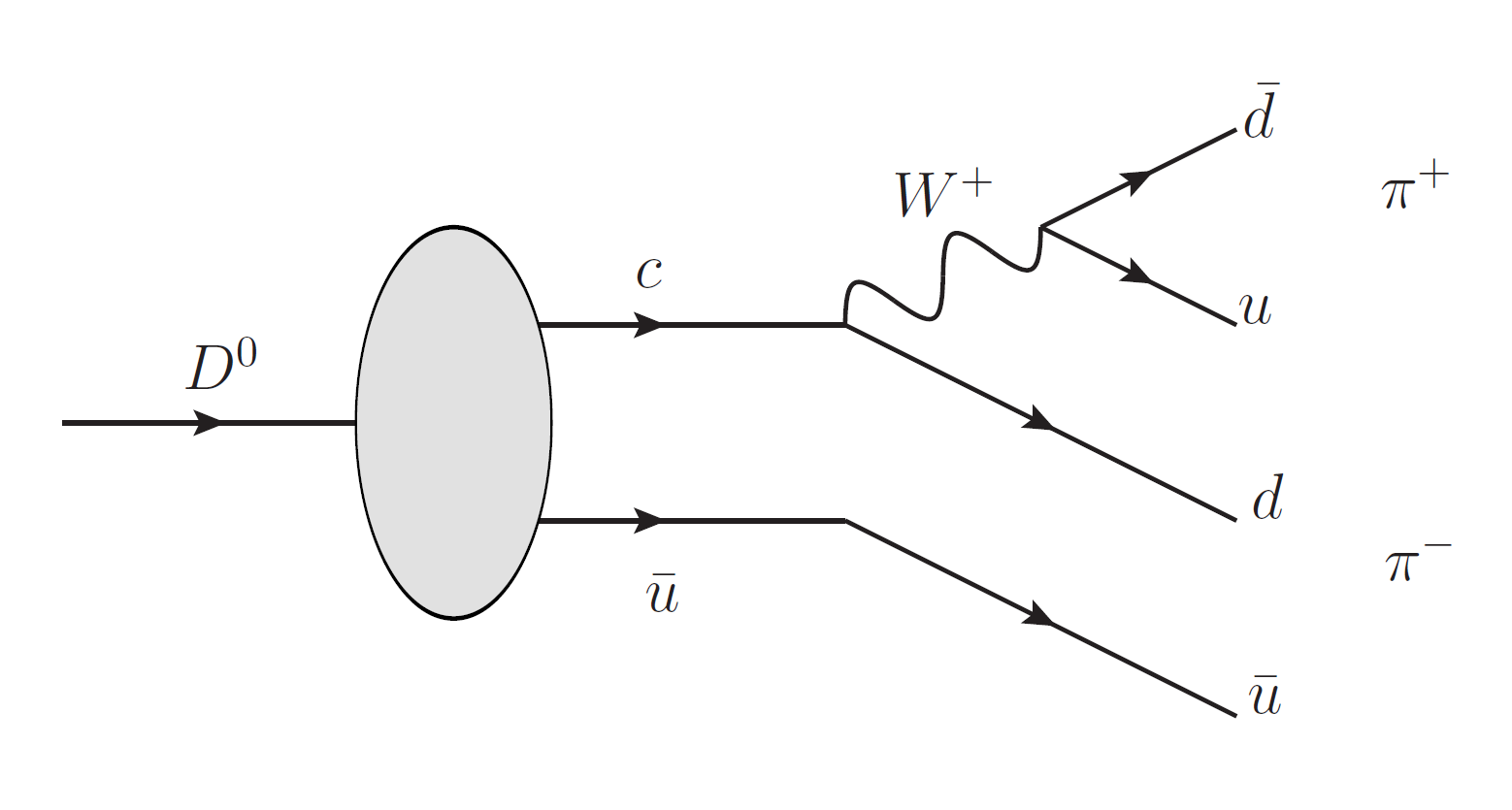}
        \caption{$D^0\rightarrow\pi^+\pi^-$ tree diagram}
        \label{fig:D0pit}
    \end{subfigure}
    \hfill
    \begin{subfigure}[b]{0.49\textwidth}
        \includegraphics[width=0.9\linewidth]{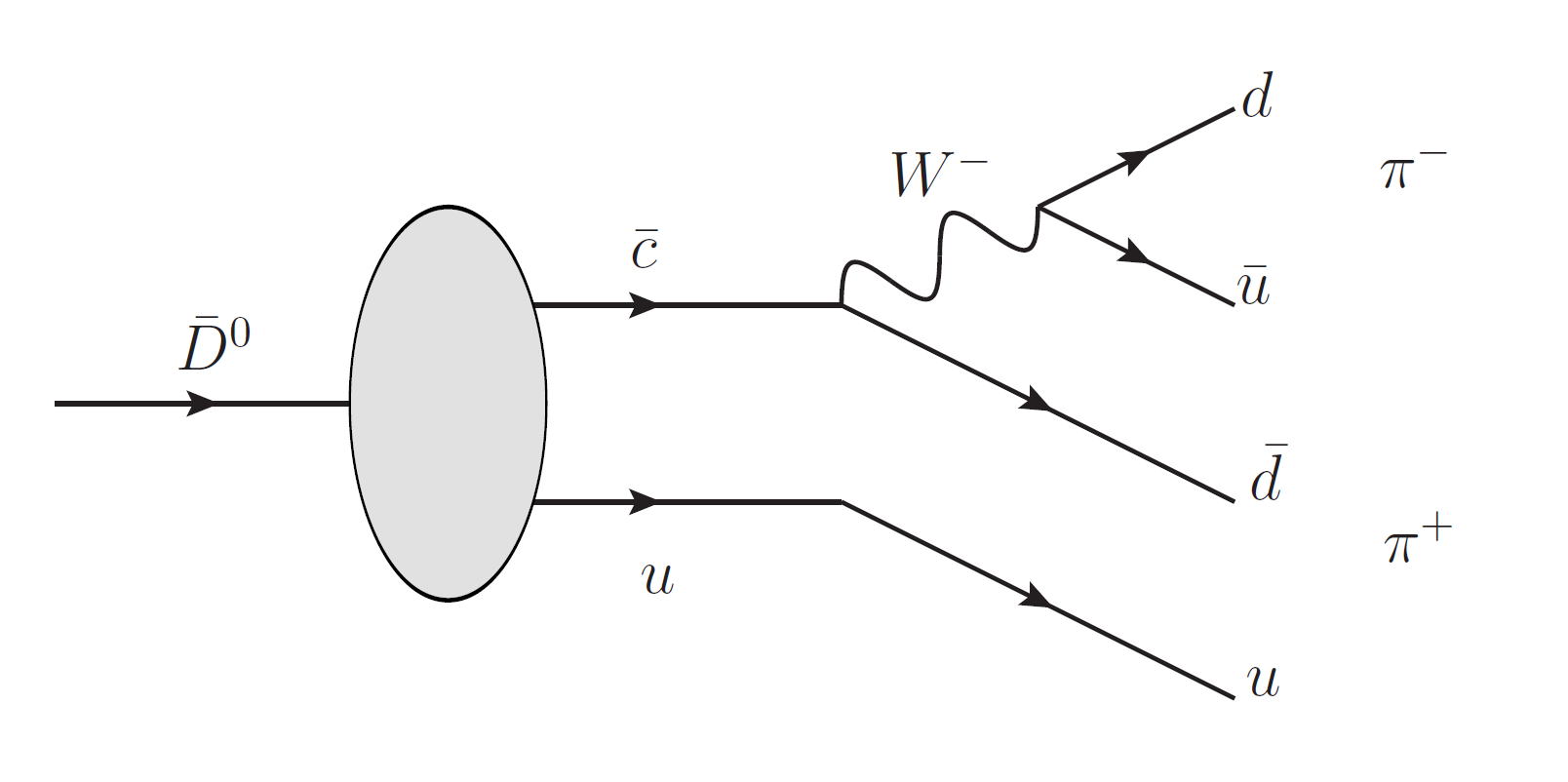}
        \caption{$\Bar{D}^0\rightarrow\pi^+\pi^-$ tree diagram}
        \label{fig:bD0pit}
    \end{subfigure}
    \vspace{-0.2cm}
    \caption{Tree diagrams for $D^0$ decays into a pair of pions}
    \label{fig:DpiT}
\end{figure}

\begin{figure}[H]
    \vspace{-0.5cm}
    \centering
    \begin{subfigure}[b]{0.49\textwidth}
        \includegraphics[width=0.9\linewidth]{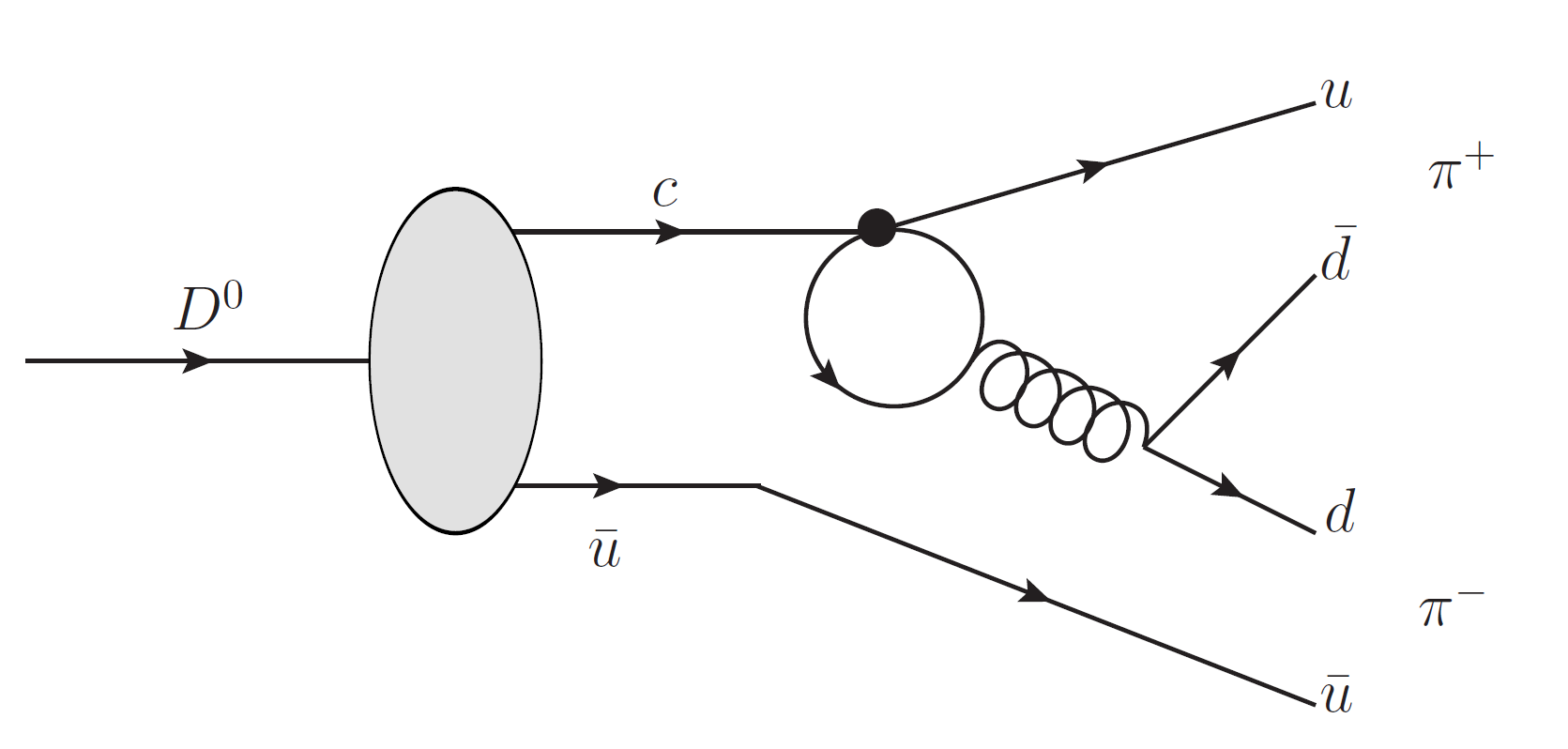}
        \caption{$D^0\rightarrow\pi^+\pi^-$ penguin diagram}
        \label{fig:D0pip}
    \end{subfigure}
    \hfill
    \begin{subfigure}[b]{0.49\textwidth}
        \includegraphics[width=0.9\linewidth]{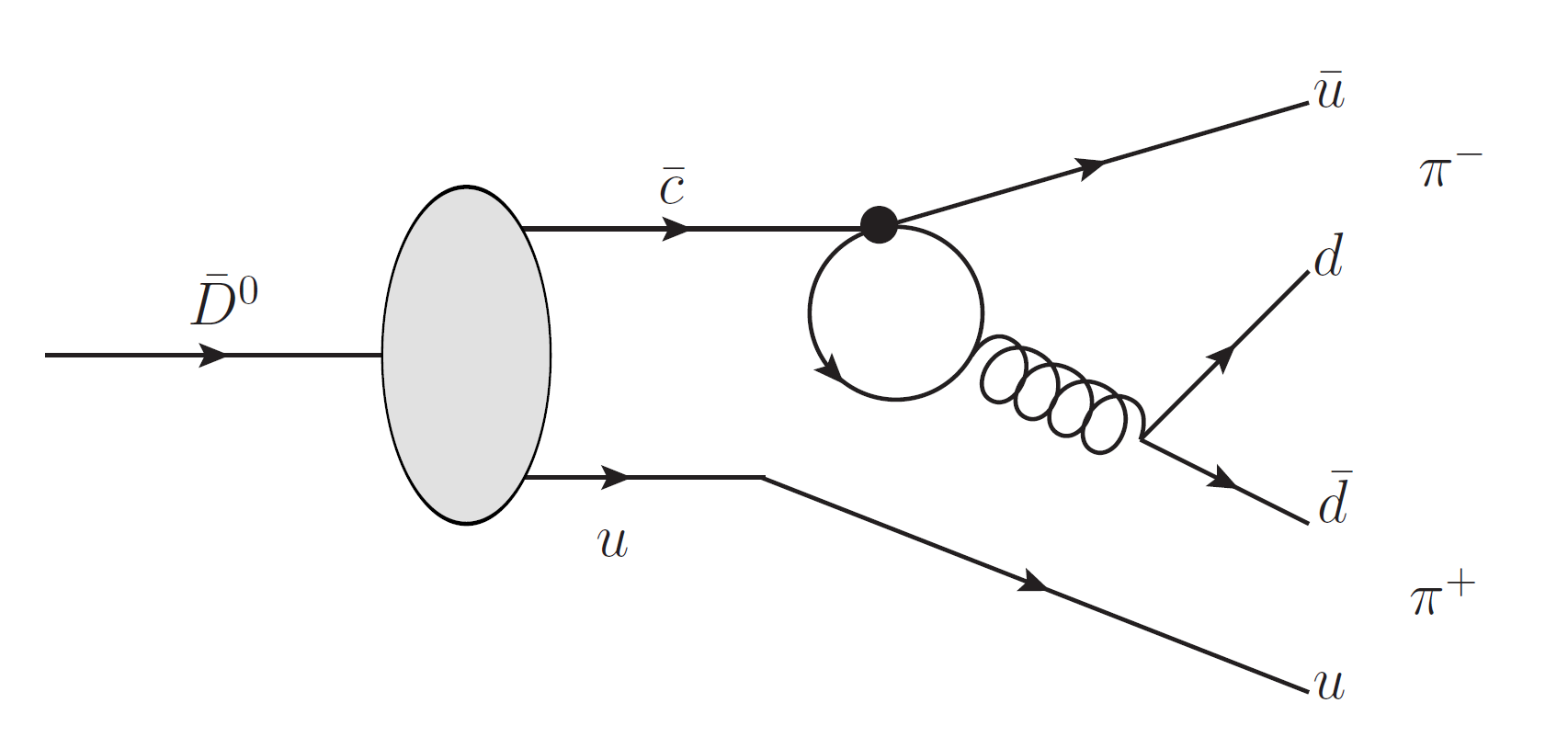}
        \caption{$\Bar{D}^0\rightarrow\pi^+\pi^-$ penguin diagram}
        \label{fig:bD0pip}
    \end{subfigure}
    \vspace{-0.2cm}
    \caption{Penguin diagrams for $D^0$ decays into a pair of pions}
    \label{fig:DpiP}
\end{figure}

\begin{figure}[H]
    \vspace{-0.5cm}
    \centering
    \begin{subfigure}[b]{0.49\textwidth}
        \includegraphics[width=0.9\linewidth]{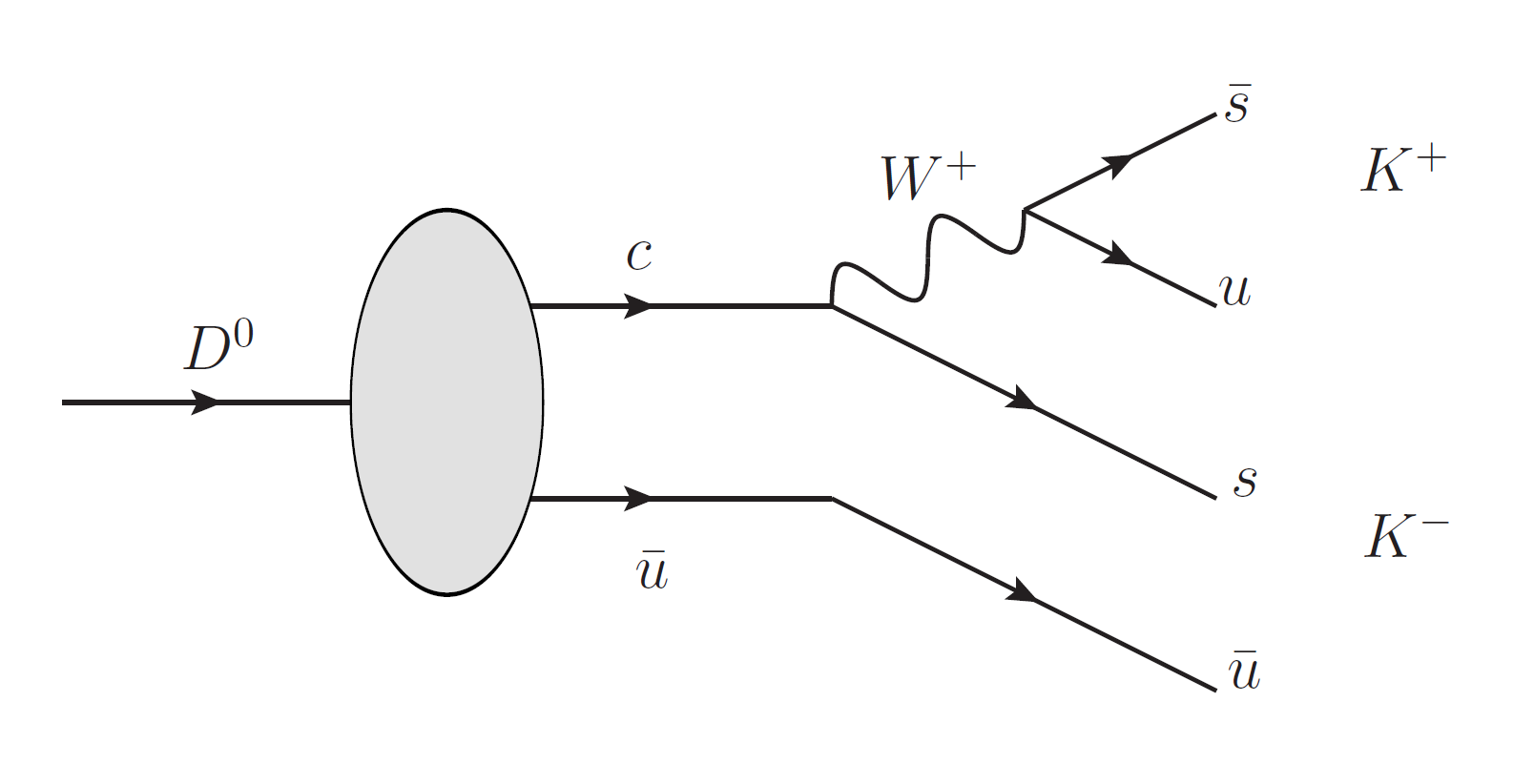}
        \caption{$D^0\rightarrow K^+K^-$ tree diagram}
        \label{fig:D0Kt}
    \end{subfigure}
    \hfill
    \begin{subfigure}[b]{0.49\textwidth}
        \includegraphics[width=0.9\linewidth]{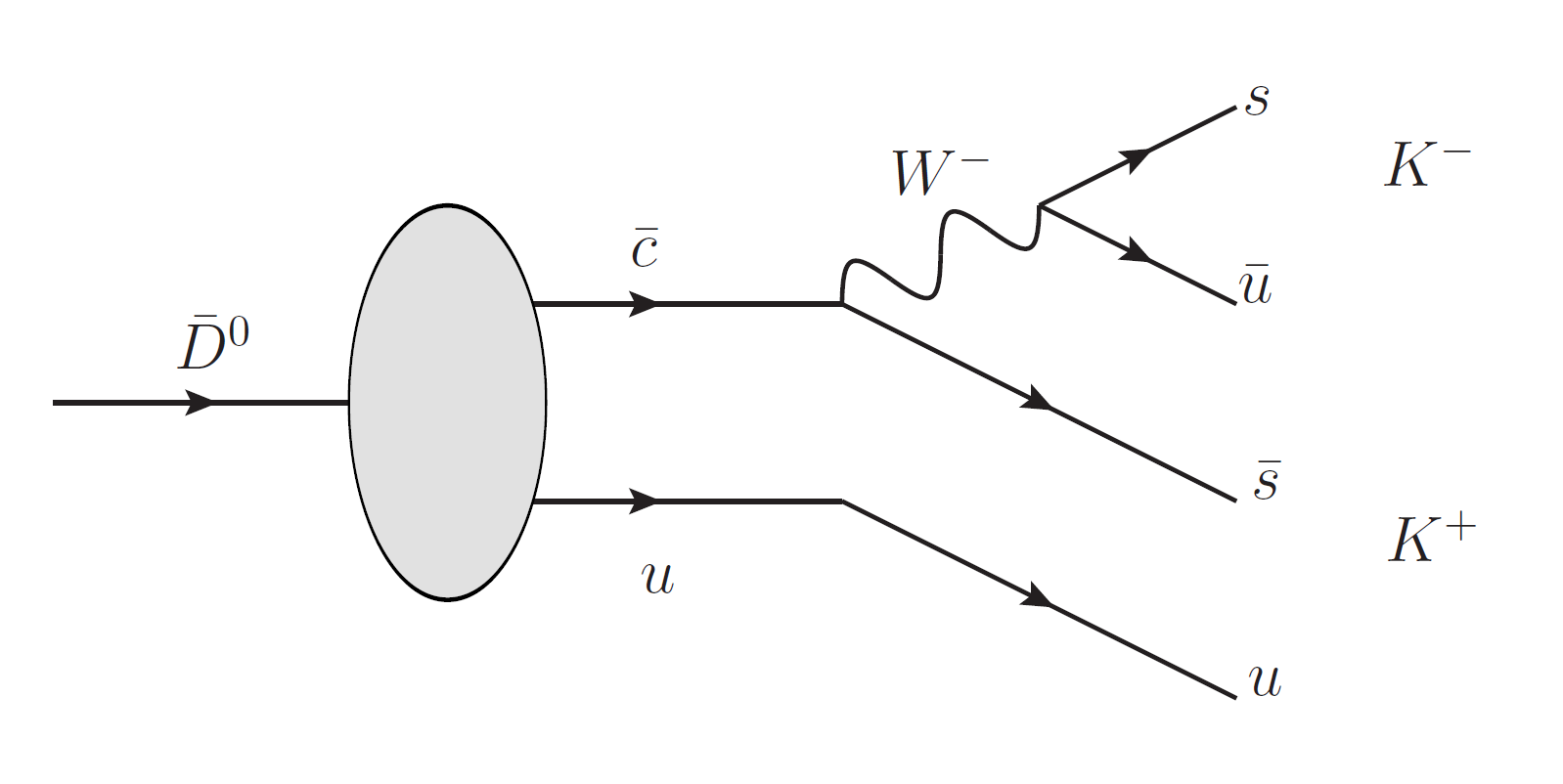}
        \caption{$\Bar{D}^0\rightarrow K^+K^-$ tree diagram}
        \label{fig:bD0Kt}
    \end{subfigure}
    \vspace{-0.2cm}
    \caption{Tree diagrams for $D^0$ decays into a pair of kaons}
    \label{fig:DKT}
\end{figure}

\begin{figure}[H]
    \vspace{-0.5cm}
    \centering
    \begin{subfigure}[b]{0.49\textwidth}
        \includegraphics[width=0.9\linewidth]{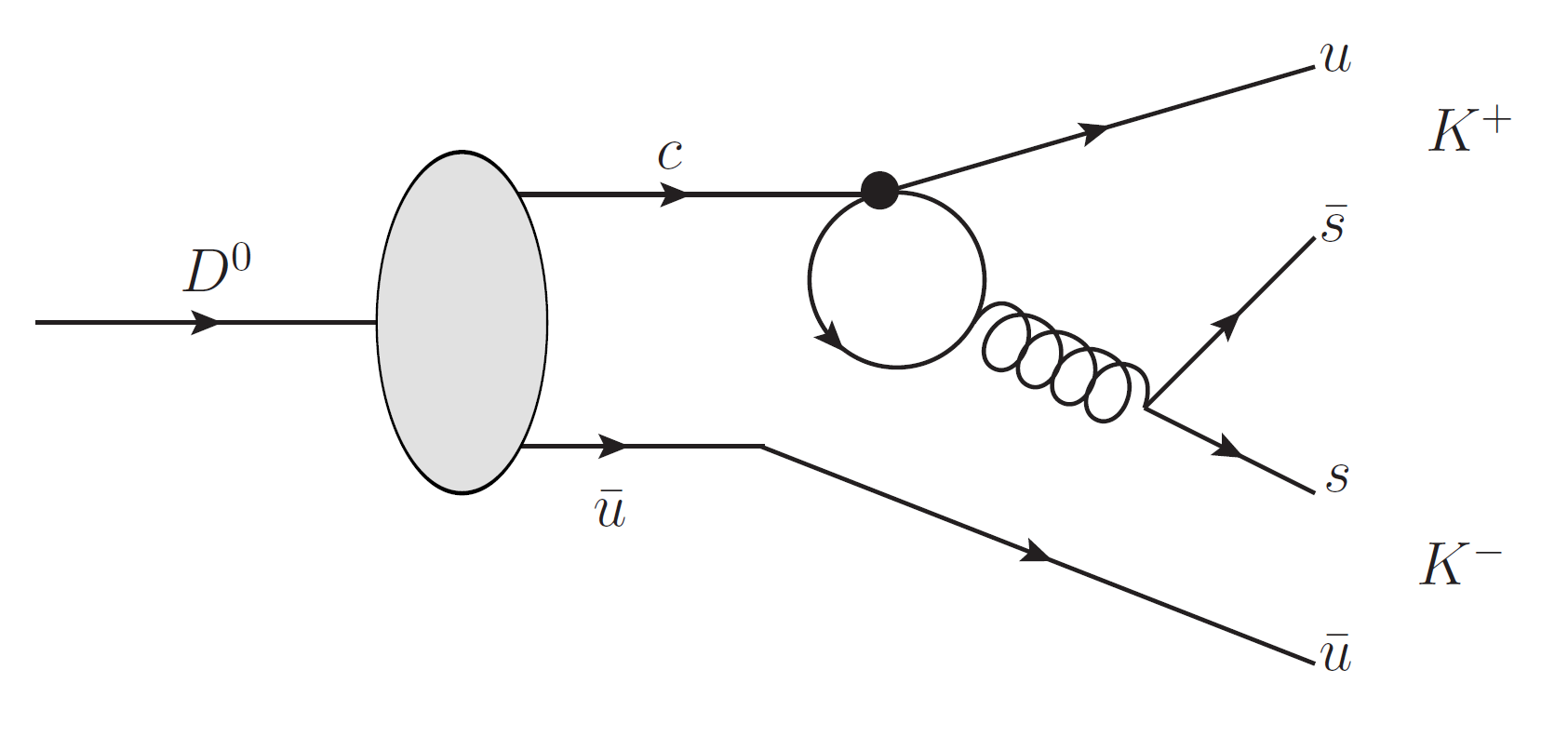}
        \caption{$D^0\rightarrow K^+K^-$ penguin diagram}
        \label{fig:D0Kp}
    \end{subfigure}
    \hfill
    \begin{subfigure}[b]{0.49\textwidth}
        \includegraphics[width=0.9\linewidth]{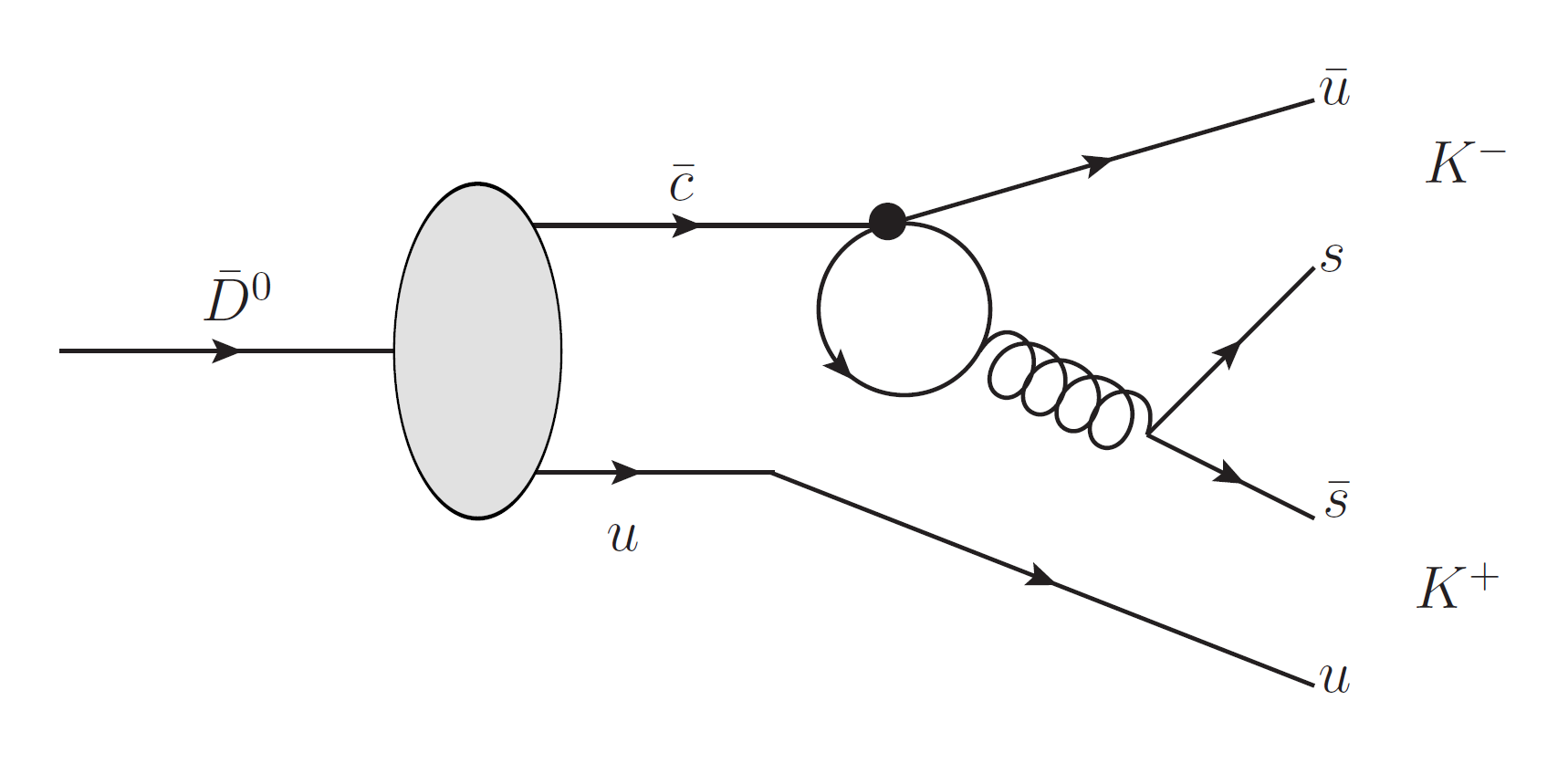}
        \caption{$\Bar{D}^0\rightarrow K^+K^-$ penguin diagram}
        \label{fig:bD0Kp}
    \end{subfigure}
    \vspace{-0.2cm}
    \caption{Penguin diagrams for $D^0$ decays into a pair of kaons}
    \label{fig:DKP}
\end{figure}

There are two types of diagrams that represent the decays: tree diagrams and penguin diagrams. They both contribute to the amplitude of the decay, but have different weighs, allowing for an interference between their amplitudes in the total process. The tree diagrams indicate direct processes of the charm quark undergoing a flavour change via a $W$ boson. On the other hand, in the penguin diagram, the charm quark creates a virtual quark via a $W$ loop and this virtual quark undergoes a strong interaction generating a quark-antiquark pair.

\subsection{CKM Matrix}

The first description of cross-generational coupling was given by Cabibbo in 1963\griffiths to explain the $u$ quark interacting with the $s$ quark and was later extended for the mixing of three quark generations changing the quark flavours. In Cabibbo's description, he defined an angle $\theta_C$, such that the interaction vertices $d \rightarrow u + W^-$ and $s \rightarrow u + W^-$ carried factors $\cos{\theta_C}$ and $\sin{\theta_C}$ respectively. With these factors, it was possible to explain the up quark coupling to the strange quark but at a smaller rate than its coupling to the down quark, since the Cabibbo angle was measured to be around $\theta_C=13.1$°. 

This description was inconsistent with some observed decay rates, such as that of the $K^0$ into a $\mu^+\mu^-$ pair. This incompleteness led to the proposition of the Glashow-Iliopoulos-Maiani (GIM) mechanism, in which a fourth quark ($c$) was introduced. The Cabibbo-GIM description allowed the interpretation that the weak eigenstates ($d'$ and $s'$, the ones which interact) are different from the mass eigenstates ($d$ and $s$)\thomson.  These weak eigenstates are linear combinations of the mass eigenstates, related by a unitary matrix, "rotating" the eigenstates by the Cabibbo angle\griffiths (equation \ref{eq:cabibbomatrix}):

\begin{equation}
    \begin{pmatrix}
    d'  \\
    s'  
    \end{pmatrix}=
    \begin{pmatrix}
    \cos{\theta_C} & \sin{\theta_C}  \\
    -\sin{\theta_C} & \cos{\theta_C}
    \end{pmatrix}\cdot
    \begin{pmatrix}
    d  \\
    s  
    \end{pmatrix}.
    \label{eq:cabibbomatrix}
\end{equation}

This mechanism can be extended to three generations\thomson of quarks, now through the $3\times3$ unitary matrix, the Cabibbo-Kobayashi-Maskawa (CKM) matrix (equation \ref{eq:CKMrotations}) relating the weak eigenstates to the mass eigenstates\thomson:

\begin{equation}
    \begin{pmatrix}
    d'  \\
    s'  \\
    b'
    \end{pmatrix}=
    \begin{pmatrix}
    V_{ud} & V_{us} & V_{ub} \\
    V_{cd} & V_{cs} & V_{cb} \\
    V_{td} & V_{ts} & V_{tb}
    \end{pmatrix}
    \begin{pmatrix}
    d  \\
    s  \\
    b
    \end{pmatrix}.
    \label{eq:CKMrotations}
\end{equation}

The CKM matrix is such that the associated vertex factor contains $V_{\alpha \beta}$ when the down, strange or bottom quarks enter the vertex as a spinor, but if one of them enters as an adjoint spinor, that is, as an antiparticle, the vertex factor is the conjugate of the CKM matrix element: $V_{\alpha \beta}^*$\thomson. These coefficients are also the determining factor of the relative strength of an interaction vertex: in the QFT description of weak interaction a weak charged-current is associated to each vertex, and the CKM element of the interacting quarks enters the expression, hence defining the interaction strength. This may cause an interaction vertex to be favored or suppressed, altering the rate at which it occurs in nature.

From the CKM matrix we interpret that each element $V_{\alpha\beta}$ describes the strength of flavour changing couplings between quarks $\alpha$ and $\beta$. Since this matrix must be unitary ($VV^{\dagger}=\mathbb{1}$, where $V$ is the CKM matrix and $\mathbb{1}$ is the identity), there are nine restrictions to its coefficients (equation \ref{eq:CKMconstraints})

\begin{equation}
    \begin{aligned}
        V_{ud}V_{ud}^*+V_{us}&V_{cd}^*+V_{ub}V_{td}^*=1 \\
        V_{ud}V_{us}^*+V_{us}&V_{cs}^*+V_{ub}V_{ts}^*=0 \\
        &...                   
    \end{aligned}
    \label{eq:CKMconstraints}
\end{equation}

Applying such constraints, the CKM matrix can be described by three rotation angles between generations ($\theta_{12}$, $\theta_{13}$ and $\theta_{23}$) and a complex phase factor ($\delta$). As seen in equation \ref{eq:CKM}, the CKM consists of the combination of the three rotation matrices. The condition of unitarity does not require it to be a real matrix, which allow for the existence of six complex phases. However, five of these phases have no physical significance and can be can be absorbed into the definition of the particles, resulting in the the introduction of the complex phase $\delta$.

\begin{equation}
\begin{aligned}
    V_{CKM}=&
    \begin{pmatrix}
    V_{ud} & V_{us} & V_{ub} \\
    V_{cd} & V_{cs} & V_{cb} \\
    V_{td} & V_{ts} & V_{tb}
    \end{pmatrix}=
    \begin{pmatrix}
    1 & 0 & 0 \\
    0 & c_{23} & s_{23} \\
    0 & -s_{23} & c_{23}
    \end{pmatrix}\times
    \begin{pmatrix}
    c_{13} & 0 & s_{13}e^{-i\delta} \\
    0 & 1 & 0 \\
    -s_{13}e^{i\delta} & 0 & c_{13}
    \end{pmatrix}\times
    \begin{pmatrix}
    c_{12} & s_{12} & 0 \\
    s_{12} & c_{12} & 0 \\
    0 & 0 & 1
    \end{pmatrix}= \\
    &=
    \begin{pmatrix}
    c_{12}c_{13} & s_{12}c_{13} & s_{13}e^{-i\delta} \\
    -s_{12}c_{23}-c_{12}s_{23}s_{13}e^{i\delta} & c_{12}c_{23}-s_{12}s_{23}s_{13}e^{i\delta} & s_{23}c_{13} \\
    s_{12}s_{23}-c_{12}c_{23}s_{13}e^{i\delta} & -c_{12}s_{23}-s_{12}c_{23}s_{13}e^{i\delta} & c_{23}c_{13}
    \end{pmatrix},
\end{aligned}
    \label{eq:CKM}
\end{equation}

where $s_{ij}-\sin{\theta_{ij}}$ and $c_{ij}-\cos{\theta_{ij}}$\thomson.

Unlike leptons, quarks do not propagate as free particles, leading to hadronisation at the length scale of $10^{-15}$ m. Hadronisation is a consequence of the asymptotic freedom of quarks, such that for the mentioned length scale, the quarks are bounded to mesons or baryons. As a consequence, the final states of their interactions must be described by mesons or baryons, which must be comprised by quarks of specific flavours. Since we can observe the quark flavour eigenstates that comprise hadronic states, it is possible to measure all CKM elements separately.

Through experimental probing and considering the unitarity conditions (equation \ref{eq:CKMconstraints}), respected within their uncertainties, the current known amplitudes of the CKM elements are\pdg

\begin{equation}
    \begin{pmatrix}
    |V_{ud}| & |V_{us}| & |V_{ub}| \\
    |V_{cd}| & |V_{cs}| & |V_{cb}| \\
    |V_{td}| & |V_{ts}| & |V_{tb}|
    \end{pmatrix}\approx
    \begin{pmatrix}
    0.97373\pm0.00031 & 0.2243\pm0.0008 & 0.00382\pm0.00020 \\
    0.221\pm0.004 & 0.975\pm0.006 & 0.0408\pm0.0014 \\
    0.0086\pm0.0002 & 0.0415\pm0.0009 & 1.014\pm0.029
    \end{pmatrix}.
\end{equation}

Observing the amplitudes of the CKM elements, it can be seen that the off-diagonal terms are substantially smaller than the diagonal terms, which implies that the rotation angles between the weak eigenstates and mass eigenstates are also small, $\theta_{12}\simeq13^{\circ}$, $\theta_{13}\simeq 0.21^{\circ}$, $\theta_{23}\simeq 2.4^{\circ}$. For such reason, the interactions from cross-generational couplings are said to be suppressed relative to same generation interactions. From the values of the rotation angles, we conclude that the most suppressed interactions are between first and third generation\thomson. 

In $D^0$ flavour changing decays, the factor determining the amplitude of the decay is $\lambda_q=V_{cq}V_{uq}^*$, and they have central values\textsuperscript{\cite{review}}

\begin{equation}
\begin{aligned}
    \lambda_{d}&=-0.21874-2.51\times 10^{-5}i\\
    \lambda_{s}&=-0.21890-0.13\times 10^{-5}i\\
    \lambda_{b}&=-1.5\times10^{-4}+2.64\times10^{-5}i.
\end{aligned}
\end{equation}

In the penguin diagram (Figures \ref{fig:DpiP} and \ref{fig:DKP}), the virtual quark can be a $d$, $s$ or $b$ quark, so the amplitude of a decay such as the $D^0\rightarrow\pi^+\pi^-$ decay can be decomposed as in equation \ref{eq:ADpi}\textsuperscript{\cite{review}}, with $A_{tree}$ and $A_{peng}$ being the amplitude contribution of each diagram. Since the tree diagram describes only the direct process with the $d$ quark, there are no contributions of the $s$ and $b$ quark CKM elements. 

\begin{equation}
    A(D^0\rightarrow\pi^+\pi^-)= \lambda_d(A_{tree}+A^d_{peng})+ \lambda_sA^s_{peng}+ \lambda_bA^b_{peng}.
    \label{eq:ADpi}
\end{equation}

A perturbative approach allows the estimate of the ratio between penguin and tree diagrams as $\vert P/T\vert\approx 0.1$\textsuperscript{\cite{review}}.

\subsubsection{Wolfenstein Parametrization}

The near-diagonal behavior of the CKM matrix can be expressed more conveniently through the Wolfenstein Parametrization. This parametrization defines four real parameters $\lambda, A, \rho$ and $\eta$ as\pdg

\begin{equation}
    \begin{aligned}
        &s_{12}=\lambda=\frac{|V_{us}|}{\sqrt{|V_{ud}|^2+|V_{us}|^2}}, \hspace{1cm} s_{23}=A\lambda^2=\lambda\left\vert \frac{V_{cb}}{V_{us}} \right\vert \\
        &s_{13}e^{i\delta}=V_{ub}^*=A\lambda^3(\rho+i\eta)=\frac{A\lambda^3(\Bar{\rho}+i\Bar{\eta})\sqrt{1-A^2\lambda^4}}{\sqrt{1-\lambda^2}(1-A^2\lambda^4(\Bar{\rho}+i\Bar{\eta}))}.
    \end{aligned}
\end{equation}

With such parametrisation, $V_{CKM}$ can be expressed as\thomson:

\begin{equation}
    V_{CKM}=
    \begin{pmatrix}
    1-\lambda^2/2 & \lambda & A\lambda^3(\rho-i\eta) \\
    -\lambda & 1-\lambda^2/2 & A\lambda^2 \\
    A\lambda^3(1-\rho-i\eta) & -A\lambda^2 & 1
    \end{pmatrix}
    +\order{\lambda^4}.
\end{equation}

As can be seen, the complex component resides in $V_{ub}$ and $V_{td}$ up to the order of $\lambda^3$, but in higher orders, proportional to $\lambda^5$, the non-zero component $\eta$ generating CP violation also appears in the elements $V_{cd}$ and $V_{ts}$. If $\eta$ is non-zero, the CKM matrix is certain to have an irreducible complex phase, which is an essential condition for CP to be violated in the quark sector\thomson. The central values of CKM elements involving the charm quark show this difference in order of magnitude in the complex phase, being around\textsuperscript{\cite{review}}

\begin{equation}
\begin{aligned}
    V_{cd}&=-0.2245-2.6\times 10^{-5}i\\
    V_{cs}&=-0.97359-5.9\times 10^{-6}i\\
    V_{cb}&=0.0416.
\end{aligned}
\end{equation}

Given the definition of $\lambda$ and the known values for the rotation angle $\theta_{12}$, it is known that $\lambda \approx 0.226$\pdg. This value allows the clear visualization that the higher $\lambda$ order attributed to one of the matrix elements, the smaller it is by nearly one order of magnitude. Hence, the emergence of the complex phase exclusively in higher orders of $\lambda$ in the $V_{cd}$ and $V_{cs}$ matrix elements allows only a small contribution of this complex phase to charm flavour changing processes. As a consequence, CP violation is harder to be probed in D-meson decays, as evidenced by the earlier discovery in B-meson decays.

\section{CP Violation}

With the general understanding of the dynamics responsible for charged-current weak interactions between quarks, the phenomenon of Charge-Parity violation can be explained. Given a certain physical process, if it is identical under charge conjugation and parity transformations, it respects Charge-Parity symmetry, however, if the dynamics of the process are altered after the transformations, it violates CP.

Parity transformation is responsible for inverting the space coordinate of a system: a coordinate $\Vec{x}$ goes to $-\Vec{x}$\bigi. This operation is responsible for determining two types of vectors and two types of scalars under rotations: polar vectors and axial vectors as well as scalars and pseudoscalars. Different observables may fall under different categories. 

Under parity transformations, polar vectors undergo a change in sign ($\Vec{V}\xrightarrow{\textbf{P}}-\Vec{V}$), meanwhile axial vectors remain unchanged ($\Vec{A}\xrightarrow{\textbf{P}}\Vec{A}$)\bigi. Scalars also remain unchanged ($S\xrightarrow{\textbf{P}} S$) and pseudoscalars go through the inversion ($P\xrightarrow{\textbf{P}}-P$)\bigi.

Charge conjugation is responsible for transforming a particle into its own antiparticle\bigi. That is, the "original" particle goes to a particle with equal mass, momentum and spin, but opposite quantum numbers, such as isospin or electric charge\bigi. As previously discussed, the existence of the antiparticle is a necessary consequence of the Dirac equation when discussing relativistic  quantum mechanics. The consequence of this inversion for the particle`s interactions is that the antiparticles enter an interaction vertex as the adjoint spinor.

If a system undergoes both a charge conjugation and a parity transformation, then the resulting system is called its CP conjugate.

Another important transformation we must acknowledge when discussing CP symmetry is time reversal transformation. It is responsible for an inversion of the time coordinate, from $t$ into $-t$\bigi. Its relevance arises from the observed CPT symmetry in our universe.

\subsection{CPT Theorem}

In order to discuss charge conjugation, parity and time reversal symmetries within the context of elementary particles, a description of discrete symmetries in the framework of the quantum field theories responsible for describing particle interactions is needed\bigi.

In quantum field theories, Lagrangians can be used to describe particle interactions. And, when analysing the effect of the mentioned CP transformations (equation \ref{eq:CPlagr}) on a general interaction Lagrangian (equation \ref{eq:lagr}), it is possible to observe that CP symmetry is conserved if the parameter $c$ is real\bigi.

\begin{equation}
\begin{aligned}
    \mathcal{L}_{T}=&aV_{\mu}^+(t, \Vec{x})V^{\mu,-}(t, \Vec{x})+bA_{\mu}^+(t, \Vec{x})A^{\mu,-}(t, \Vec{x}) \\
    &+cV_{\mu}^+(t, \Vec{x})A^{\mu,-}(t, \Vec{x})+c^*A_{\mu}^+(t, \Vec{x})V^{\mu,-}(t, \Vec{x})
    \label{eq:lagr}
\end{aligned}
\end{equation}

\begin{equation}
\begin{aligned}
    \mathbf{CP}\mathcal{L}_{T}\mathbf{CP}^{\dagger}=&aV^{\mu,-}(t, -\Vec{x})V_{\mu}^+(t, -\Vec{x})+bA^{\mu,-}(t, -\Vec{x})A_{\mu}^+(t, -\Vec{x}) \\
    &+cV^{\mu,-}(t, -\Vec{x})A_{\mu}^+(t, -\Vec{x})+c^*A^{\mu,-}(t, -\Vec{x})V_{\mu}^+(t, -\Vec{x})
    \label{eq:CPlagr}
\end{aligned}
\end{equation}

However, the most manageable observable in this context is the mass of a particle, directly related to its energy. For such reason, it is adequate to study the problem under Hamiltonian formalism\bigi. Within this framework, the Hamiltonian can be expressed in terms of local operators (equation \ref{eq:localh}), where $\mathbf{CP}H_i\mathbf{CP}^{\dagger}=H_i^{\dagger}$. Therefore, CP is conserved if $a_i$ in equation \ref{eq:localh} is a real parameter\bigi. 

\begin{equation}
    H=\sum_i a_iH_i+h.c.
    \label{eq:localh}
\end{equation}

This local description, uniting the demands of quantum mechanics and special relativity requires the existence of antiparticles\bigi. However, applying the same reasoning with the CPT operator, it is possible to postulate the CPT theorem: the combined transformation CPT can always be defined - as an anti-unitary operator - in such a way for a local quantum field theory that it represents a symmetry: 

\begin{equation}
    \mathbf{CPT}\mathcal{L}(t,\Vec{x})(\mathbf{CPT})^{-1}=\mathcal{L}(-t,-\Vec{x}).
\end{equation}

This theorem can be proven by assuming (a) Lorentz invariance, (b) the existence of a unique vacuum state and (c) weak local commutativity obeying the right statistics\bigi.

Through the transformation of the Lagrangian, CPT is always conserved, independent of the nature of the coupling parameters $a$, $b$ and $c$. The construction of this argument was done by defining bosons in terms of fermionic bilinears, thus it can be extended for fermions under a similar assumption\bigi.

For our discussion, the consequences of the CPT theorem are far more important than the deduction of the theorem itself. The most important ones are the equality of masses and lifetimes for particles and antiparticles.

The mass of a particle $M(P)$ can be proven to be equal to the mass of its antiparticle $M(\bar{P})$\bigi:

\begin{equation}
    \begin{aligned}
        M(P)=&\expval{H}{P}
        =\expval{(\mathbf{CPT})^{\dagger}\mathbf{CPT}H(\mathbf{CPT})^{-1}\mathbf{CPT}}{P}^* \\
        =&\expval{\mathbf{CPT}H(\mathbf{CPT})^{-1}}{\bar{P}}^*
        =\expval{H}{\bar{P}}^*=M(\Bar{P}),
    \end{aligned}
\end{equation}

where H is the Hamiltonian, and $P$($\Bar{P}$) is the particle(antiparticle) eigenstate.

Similarly, we obtain the equivalence of lifetimes $\Gamma(P)=\Gamma(\Bar{P})$: 

\begin{equation}
    \begin{aligned}
        \Gamma(P)=&2\pi\sum_{f}\delta(M_P-E_f)\vert \mel{f;out}{H_{decay}}{P} \vert^2 \\
        =&2\pi\sum_{f}\delta(M_P-E_f)\vert \mel{f;out}{(\mathbf{CPT})^{\dagger}\mathbf{CPT}H_{decay}(\mathbf{CPT})^{-1}\mathbf{CPT}}{P}^* \vert^2 \\
        =&2\pi\sum_{f}\delta(M_P-E_f)\vert \mel{\Bar{f};in}{H_{decay}}{\Bar{P}} \vert^2 \\
        =&2\pi\sum_{f}\delta(M_P-E_f)\vert \mel{\Bar{f};out}{H_{decay}}{\Bar{P}} \vert^2 \\
        =&\Gamma(\Bar{P}),
    \end{aligned}
\end{equation}

where we used the properties that, under time reversal, multi-particle states transform as in equation \ref{eq:MPtime} and that both "in" and "out" states form complete sets of states (equation \ref{eq:inout})\bigi.

\begin{equation}
    \mathbf{T}\ket{p_1, p_2, ...; out}=\ket{-p_1, -p_2, ...; in},
    \label{eq:MPtime}
\end{equation}

\begin{equation}
    \sum_{f}\ket{f; in}\bra{f; in}=\sum_{f}\ket{f; out}\bra{f; out}=1.
    \label{eq:inout}
\end{equation}

The equality of lifetimes, along with the completeness condition\bigi may be further used to conclude that classes $F_i$ containing final states $f_{\alpha}^{(i)}$ mutually distinct under strong interactions can be used to separate decay channels. As an example, take final states $f_{\alpha}^{(i)}\in F_i$ and $f_{\rho}^{(j)}\in F_j$ with $i\neq j$; in this case, neither $f_{\alpha}^{(i)}\rightarrow f_{\rho}^{(j)}$ nor $f_{\rho}^{(j)}\rightarrow f_{\alpha}^{(i)}$ can happen via the strong interaction, independently of $\alpha$ and $\rho$. Then, as a consequence of CPT invariance, 

\begin{equation}
    \sum_{f_{\alpha}^{(i)}\in F_i}\Gamma(P\rightarrow f_{\alpha}^{(i)})= \sum_{\Bar{f}_{\alpha}^{(i)}\in \Bar{F}_i}\Gamma(\Bar{P}\rightarrow \Bar{f}_{\alpha}^{(i)}),
\end{equation}

where $f_{\alpha}^{(i)}$ and $F_i$ are CP conjugate to $\Bar{f}_{\alpha}^{(i)}$ and $\Bar{F}_i$\bigi, respectively.

Hence, considering strong final state interactions, "in" and "out" states cannot be distinguished. In this situation, CPT implies more than just the equality of the total sum of particle decays between particles and antiparticles\bigi: the sums over certain subsets of all decay rates also have to be the same for particles and antiparticles. However, when final states are their own CP conjugates (as in the case of $\pi^+\pi^-$ and $K^+ K^-$), the presence of strong final state interactions cannot be ignored and CPT invariance does not guarantee the equality of decay rates such as $\Gamma(D^0\rightarrow\pi^+\pi^-)$ and $\Gamma(\Bar{D}^0\rightarrow\pi^+\pi^-)$.

\subsection{Final state interactions (FSI)}

The appearance of a CP asymmetry can then be associated to a difference in partial decay widths into CP eigenstates such as $\pi^+\pi^-$ or $K^+ K^-$. This asymmetry can be quantified as 

\begin{equation}
    A_{CP}\equiv \frac{\Gamma(D^0\rightarrow f)-\Gamma(\Bar{D}^0\rightarrow \Bar{f})}{\Gamma(D^0\rightarrow f)+\Gamma(\Bar{D}^0\rightarrow \Bar{f})}.
    \label{eq:acp}
\end{equation}

The partial decay widths allow for the observation of CP violation, but final state interactions are essential to explain the reason for this. A weak decay such as $D^0\rightarrow f$ can have contributions from two coherent processes of different amplitudes. This could be, for example, related to different contributions from penguin and tree diagrams. The total transition amplitude is then 

\begin{equation}
    A(D^0\rightarrow f)=e^{i\phi_{1}^{weak}}e^{i\delta_{1}^{FSI}}\vert\mathcal{A}_1\vert+ e^{i\phi_{2}^{weak}}e^{i\delta_{2}^{FSI}}\vert\mathcal{A}_2\vert,
    \label{eq:ampfsi}
\end{equation}

where $\phi_{1,2}^{weak}$ are phases due to the weak decay dynamics, and $\delta_{1,2}^{FSI}$ are due to strong final state interactions\bigi. The decay amplitude of the CP conjugate process is 

\begin{equation}
    A(\Bar{D}^0\rightarrow \Bar{f})= e^{-i\phi_{1}^{weak}}e^{i\delta_{1}^{FSI}}\vert\mathcal{A}_1\vert+ e^{-i\phi_{2}^{weak}}e^{i\delta_{2}^{FSI}}\vert\mathcal{A}_2\vert,
    \label{eq:antiampfsi}
\end{equation}

since the weak phase is opposite in sign for the antiparticle decay.

Hence, the CP asymmetry arises from this difference and, substituting in equation \ref{eq:acp} the decay rates obtained from the decay amplitudes in equations \ref{eq:ampfsi} and \ref{eq:antiampfsi}, 

\begin{equation}
    A_{CP}=-\frac{2\sin{(\Delta\phi_W)}\sin{(\Delta\delta^{FSI})}\vert\mathcal{A}_2/\mathcal{A}_1\vert}{1+\vert\mathcal{A}_2/\mathcal{A}_1\vert^2+2\vert\mathcal{A}_2/\mathcal{A}_1\vert\cos{(\Delta\phi_W)}\cos{(\Delta\delta^{FSI})}},
    \label{eq:FSIacp}
\end{equation}

where $\Delta\phi_W=\phi_1^{weak}-\phi_2^{weak}$ and $\Delta\delta^{FSI}=\phi_1^{FSI}-\phi_2^{FSI}$\bigi. 

Thus, for the CP asymmetry to arise, it is necessary that it enters through the weak dynamics, having $\Delta\phi_W\neq0$ and that strong final state interactions induce a non-trivial phase shift ($\Delta\delta^{FSI}\neq0$). The phase shifts due to FSI are the same for a given process and its CP conjugate, however, the weak phases enter those decays with opposite signs. As a result, the FSI phases can induce a T-odd correlation even with vanishing weak phases\bigi and, by comparing this same correlation in CP conjugate states, this effect can be disentangled. From equation \ref{eq:FSIacp}, it is evident that the observation of CP violation in the asymmetry of partial widths is only possible because of these FSI phase shifts, in spite of their values being unknown. This interference between the two different decay processes is what causes the CP violation. 

\subsection{Neutral Meson Oscillations}

This description of CP violation is not sufficient for analysing some neutral particle systems. Unlike neutral pions, the quark and antiquark that compose the D-meson are not their own antiparticle. For such reason, we can attribute a "charm" number of $+1$ to the $D^0$ and $-1$ to the $\bar{D}^0$. The difference in their quark composition result in them being two different particles, whose existences must be independent.

When a particle-antiparticle pair does not share this internal quantum number ("charm" in the case of the $D^0$ meson), but have no electric charge, weak and the so-called "superweak" interactions can cause a particle-antiparticle transition where the internal quantum number changes by two units.
We differ the $D^0$ and $\Bar{D}^0$ only by their internal charm quantum number, allowing for the $D$-meson to oscillate between these two states. This discussion is also valid for the altering of other quantum numbers, such as "beauty" in B-meson oscillations or "strangeness" in kaon oscillations, however, for a more direct approach, the $D^0$ case will be described.

\begin{figure}[H]
    \centering
    \hfill
    \begin{subfigure}[t]{0.48\textwidth}
        \includegraphics[width=0.95\linewidth]{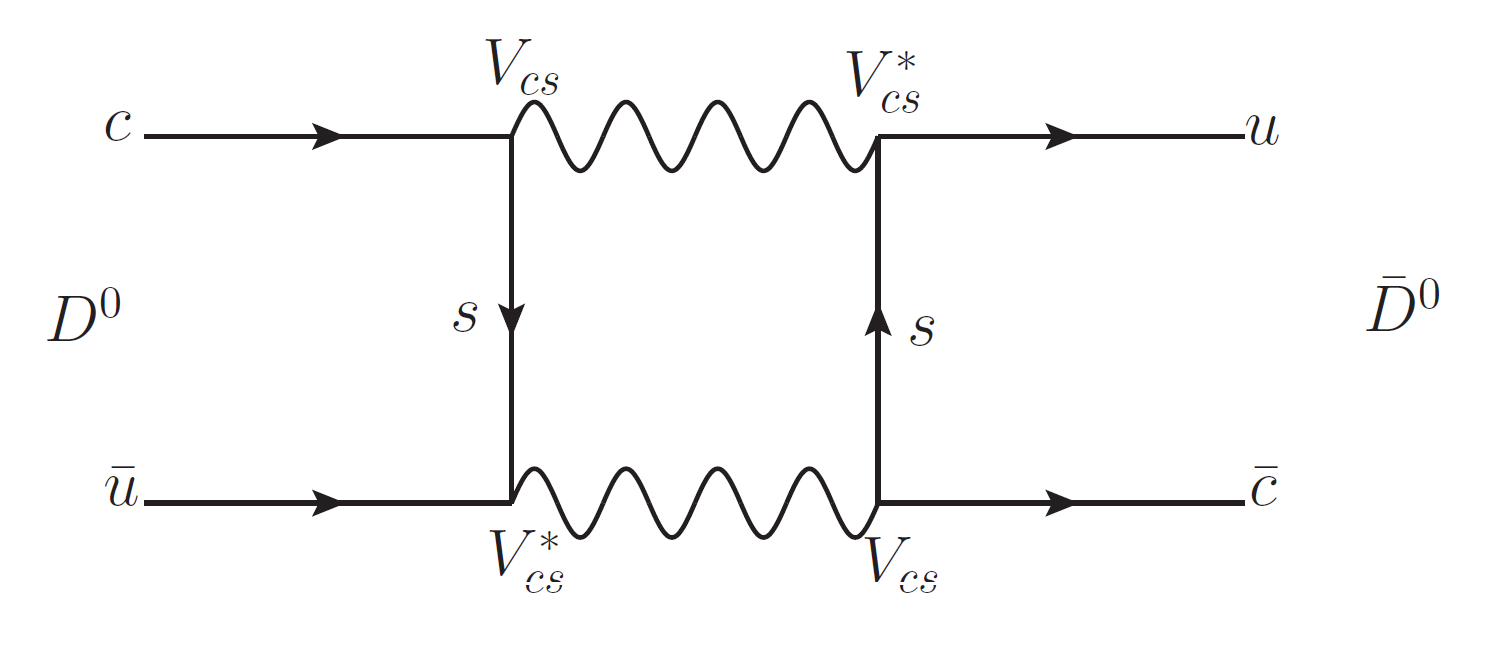}
    \end{subfigure}
    \begin{subfigure}[b]{0.48\textwidth}
        \includegraphics[width=0.95\linewidth]{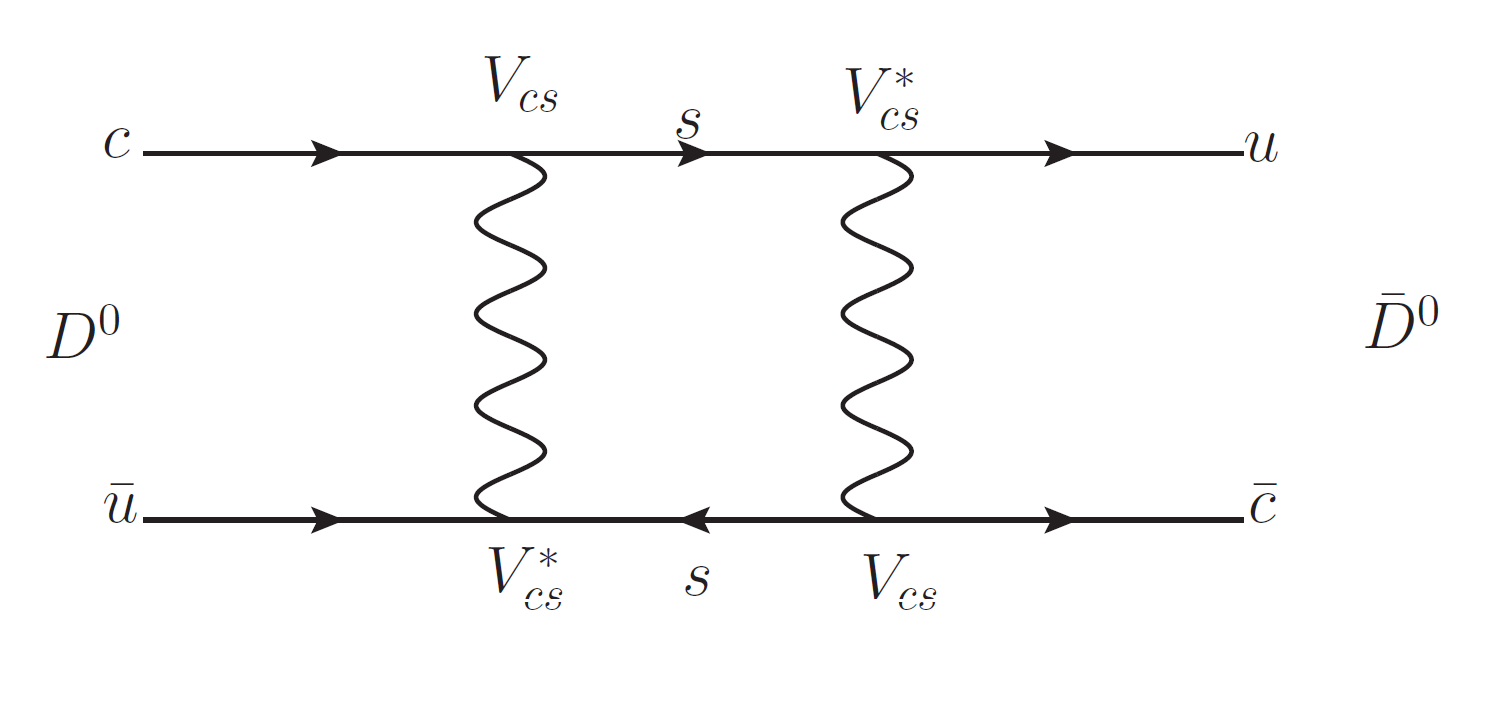}
    \end{subfigure}
    \vspace{-0.2cm}
    \caption{$D^0\leftrightarrow \Bar{D}^0$ oscillation diagrams}
    \label{fig:feynDD}
\end{figure}

This oscillation (also called mixing) can be represented by Feynman diagrams such as the ones in Figure \ref{fig:feynDD}. These diagrams display exclusively the virtual strange quarks, which have the most substantial contribution to the mixing (due to the charm quark coupling being strongest with the strange quark, mathematically equivalent to $V_{cs}>>V_{cd}$ and $V_{cs}>>V_{cb}$), but all combinations of virtual down, strange and beauty quarks are allowed.

Denoting $C$ as the charm quantum number and $H_{weak}$ as the Hamiltonian of a weak interaction, the following properties are postulated: $\Delta C=0$ for strong and electromagnetic interactions, while $\Delta C\neq0$ for weak interactions. The weak Hamiltonian can couple both $D^0$ and $\bar{D}^0$ to a common state I with $C=0$. The transitions guided by $H_{weak}$ from a $D^0$ or $\Bar{D}^0$ state into I only change $C$ by one unit, being called a weak transition. However, $H_{weak}$ can couple $D^0$ directly to $\Bar{D}^0$, thus allowing superweak forces to generate $\Delta C=2$ transitions through $D^0\leftrightarrow \Bar{D}^0$.

\subsubsection{Oscillation eigenstates}

All possible transitions can be accounted for by the complete Hamiltonian $H$, separated by these different transitions as in equation \ref{eq:Hfull},

\begin{equation}
    H=H_{\Delta C=0}+H_{\Delta C=1}+H_{\Delta C=2},
    \label{eq:Hfull}
\end{equation}

where $H_{\Delta C=0}$ contains the strong and electromagnetic forces\bigi, while $H_{\Delta C=1}$ and $H_{\Delta C=2}$ account for the weak force. 

The time evolution of the $D^0 \leftrightarrow \bar{D}^0$ system can be represented as:

\begin{equation}
    \ket{\Tilde{\Psi}(t)}=a(t)\ket{D^0}+b(t)\ket{\Bar{D}^0}+c(t)\ket{\pi \pi}+d(t)\ket{KK}+e(t)\ket{\pi l \Bar{\nu}_l}+...,
\end{equation}

where $l$ can be an electron or a muon. The coefficients of this time evolution can be obtained from the Schrödinger equation: 

\begin{equation}
    i\hbar\pdv{\Tilde{\Psi}}{t}=H\Tilde{\Psi},
    \label{eq:schro}
\end{equation}

where the Hamiltonian H is an infinite-dimensional Hermitian matrix in the Hilbert space\bigi. The incapacity of obtaining this time evolution analytically can be overcome through some considerations: the initial state is a linear combination of $D^0$ and $\bar{D}^0$: $\ket{\Psi(0)}=a(0)\ket{D^0}+b(0)\ket{\Bar{D}^0}$; since the goal is to describe the oscillations between the two states, the only coefficients of interest are $a(t)$ and $b(t)$; and the Weisskopf-Wiegner approximation, stating that the time evolution is considered only for times larger than the time scale of strong interactions.

Applying such simplifications, the dynamics of the oscillation can be described by the time evolution:

\begin{equation}
    i\hbar\pdv{\Tilde{\Psi}}{t}=\mathcal{H}\Tilde{\Psi},
\end{equation}

in which $\Psi(t)$ is restricted to the subspace of $D^0$ and 
$\Bar{D}^0$:

\begin{equation}
    \Psi(t)=
    \begin{pmatrix}
        a(t) \\
        b(t)
    \end{pmatrix},
\end{equation}

and the Hamiltonian operator matrix is given by:

\begin{equation}
    \mathcal{H}=\mathbf{M}-\frac{i}{2}\boldsymbol{\Gamma}=
    \begin{pmatrix}
        M_{11}-\frac{i}{2}\Gamma_{11} & M_{12}-\frac{i}{2}\Gamma_{12} \\
        M_{21}-\frac{i}{2}\Gamma_{21} & M_{22}-\frac{i}{2}\Gamma_{22}
    \end{pmatrix}.
    \label{eq:H}
\end{equation}.

This is a perturbative description, assuming that oscillation is a consequence of non-zero off-diagonal terms, which would otherwise be diagonal if the mass eigenstates remained unaltered.

Assuming that H commutes with C, P and T, these operators can be defined as

\begin{equation}
    \mathbf{C}\ket{D^0}=-\ket{\Bar{D}^0},\hspace{1cm} \mathbf{P}\ket{D^0}=-\ket{D^0},\hspace{1cm} \mathbf{T}\ket{D^0}=\ket{D^0}, 
\end{equation}

such that 

\begin{equation}
    \begin{aligned}
        \mathbf{CP}\ket{n; out(in)}=&\ket{\Bar{n}; out(in)} \\
        \mathbf{T}\ket{n; in(out)}=&\ket{n; out(in)}.
    \end{aligned}
\end{equation}

The "in" and "out" states can be assumed to form equivalent complete bases, from which it can be concluded that $M_{11}=M_{22}$ and $\Gamma_{11}=\Gamma_{22}$ through CPT and CP symmetry, as well as $M_{12}=M_{21}$ and $\Gamma_{12}=\Gamma_{21}$ from CP and T symmetry\bigi:

\begin{equation}
    \begin{aligned}
        \mathbf{CPT}\text{ or }\mathbf{CP}\text{ invariance} &\implies M_{11}=M_{22}, \Gamma_{11}=\Gamma_{22} \\
        \mathbf{CP}\text{ or }\mathbf{T}\text{ invariance} &\implies \Im{M_{12}}=\Im{\Gamma_{12}}=0
    \end{aligned}
    \label{eq:MGConstraints}
\end{equation}

\subsubsection{$\mathcal{H}$ diagonalization into mass eigenstates}

The diagonalization of the $\mathcal{H}$ matrix yields two decoupled equations, each having as solution one of the two mass eigenstates, $\ket{D_1}$ and $\ket{D_2}$. Since $\mathbf{M}$ and $\boldsymbol{\Gamma}$ are hermitian, they can be diagonalized by unitary transformations. The first step is to express $\mathcal{H}$ in terms of the Pauli matrices and the identity

\begin{equation}
    \mathcal{H}=\mathbf{M}-\frac{i}{2}\boldsymbol{\Gamma}=E_1\sigma_1+E_2\sigma_2+E_3\sigma_3-iD\mathbb{1},
    \label{eq:MGpauli}
\end{equation}

where the Pauli matrices are defined as

\begin{equation}
    \sigma_1=
    \begin{pmatrix}
        0 & 1 \\
        1 & 0
    \end{pmatrix},
    \sigma_2=
    \begin{pmatrix}
        0 & -i \\
        i & 0
    \end{pmatrix},
    \sigma_3=
    \begin{pmatrix}
        1 & 0 \\
        0 & -1
    \end{pmatrix}.
\end{equation}

From equation \ref{eq:MGpauli} the following relations arise:

\begin{equation}
    \begin{aligned}
        E_1&=\Re\{M_{12}\}-\frac{i}{2}\Re\{\Gamma_{12}\} \\
        E_2&=-\Im\{M_{12}\}+\frac{i}{2}\Im\{\Gamma_{12}\} \\
        E_3&=\frac{1}{2}(M_{11}-M_{22})-\frac{i}{4}(\Gamma_{11}-\Gamma_{22}) \\
        D&=\frac{i}{2}(M_{11}+M_{22})+\frac{1}{4}(\Gamma_{11}+\Gamma_{22}).
    \end{aligned}
\end{equation}

Introducing spherical coordinates allows for a more concise form of expressing the diagonalized mass eigenstates. Hence, the complex coordinates $E$, $\theta$, $\phi$ are defined, such that\bigi:

\begin{equation}
    \begin{aligned}
        E=&\sqrt{E_1^2+E_2^2+E_3^2} \\
        E_1= E\sin{\theta}\cos{\phi}, \hspace{1cm} E_2&= E\sin{\theta}\sin{\phi}, \hspace{1cm}E_1= E\cos{\theta}.
    \end{aligned}
\end{equation}

The complex coordinates require the usual definition for sine and cosine functions\bigi:

\begin{equation}
    \cos{z}=\frac{1}{2}\left(e^{iz}+e^{-iz}\right), \hspace{1cm} \sin{z}=\frac{1}{2i}\left(e^{iz}-e^{-iz}\right).
\end{equation}

In order for $D^0\leftrightarrow\Bar{D}^0$ oscillations to occur, $\mathbf{M}-\frac{1}{2}\boldsymbol{\Gamma}$ is not originally diagonal, therefore the condition for these oscillations are\bigi:

\begin{equation}
    E\neq0, \hspace{1cm} \sin{\theta}\neq0.
\end{equation}

The previous symmetry constraints on the elements of the $\mathbf{M}$ and $\boldsymbol{\Gamma}$ matrices (equation \ref{eq:MGConstraints}) can be expressed in terms of the complex coordinates:

\begin{equation}
    \begin{aligned}
        \mathbf{CPT}\text{ or }\mathbf{CP}\text{ invariance} &\implies \cos{\theta}=0 \\
        \mathbf{CP}\text{ or }\mathbf{T}\text{ invariance} &\implies \phi=0.
    \end{aligned}
    \label{eq:CPTconstraint}
\end{equation}

Through the defined coordinates, the mass eigenstates can be written as linear combinations of the weak eigenstates as\bigi:

\begin{equation}
    \begin{aligned}
        \ket{D_1}=&p_1\ket{D^0}+q_1\ket{\Bar{D}^0} \\
        \ket{D_2}=&p_2\ket{D^0}-q_2\ket{\Bar{D}^0}.
    \end{aligned}
\end{equation}

where the coefficients (equation \ref{eq:p1p2q1q2}) and normalization factors (equation \ref{eq:N1N2}) are

\begin{equation}
    \begin{aligned}
        p_1=N_1\cos{\frac{\theta}{2}},& \hspace{1cm} q_1=N_1e^{i\phi}\sin{\frac{\theta}{2}} \\
        p_2=N_2\sin{\frac{\theta}{2}},& \hspace{1cm} q_2=N_2e^{i\phi}\cos{\frac{\theta}{2}}
    \end{aligned}
    \label{eq:p1p2q1q2}
\end{equation}

\begin{equation}
    \begin{aligned}
        N_1=&\frac{1}{\sqrt{\left\vert \cos{\frac{\theta}{2}}\right\vert^2+\left\vert e^{i\phi}\sin{\frac{\theta}{2}}\right\vert^2}} \\
        N_2=&\frac{1}{\sqrt{\left\vert \sin{\frac{\theta}{2}}\right\vert^2+\left\vert e^{i\phi}\cos{\frac{\theta}{2}}\right\vert^2}}.
    \end{aligned}
    \label{eq:N1N2}
\end{equation}

Imposing the previously discussed and necessary CPT invariance, the constraint $\cos{\theta}=0$ (equation \ref{eq:CPTconstraint}) is obligatory, resulting in $p_1=p_2$ and $q_1=q_2$, which are rewritten as $p$ and $q$ respectively:

\begin{equation}
    \begin{aligned}
        \ket{D_1}=&p\ket{D^0}+q\ket{\Bar{D}^0} \\
        \ket{D_2}=&p\ket{D^0}-q\ket{\Bar{D}^0}.
    \end{aligned}
\end{equation}

These mass eigenstates have eigenvalues

\begin{equation}
    \begin{aligned}
        M_1-\frac{i}{2}\Gamma_1=&-iD+E=M_{11}-\frac{i}{2}\Gamma_{11}+\frac{q}{p}\left(M_{12}-\frac{i}{2}\Gamma_{12}\right) \\
        M_2-\frac{i}{2}\Gamma_2=&-iD-E=M_{11}-\frac{i}{2}\Gamma_{11}-\frac{q}{p}\left(M_{12}-\frac{i}{2}\Gamma_{12}\right),
    \end{aligned}
\end{equation}

where, as can be seen from equation \ref{eq:p1p2q1q2}

\begin{equation}
    \left(\frac{q}{p}\right)^2=\frac{M_{12}^*-\frac{i}{2}\Gamma_{12}^*}{M_{12}-\frac{i}{2}\Gamma_{12}},
\end{equation}

with two possible solutions, both equivalent under the exchange of the labels $D_1$ and $D_2$ of the mass eigenstates:

\begin{equation}
    \left(\frac{q}{p}\right)=\pm \sqrt{\frac{M_{12}^*-\frac{i}{2}\Gamma_{12}^*}{M_{12}-\frac{i}{2}\Gamma_{12}}}.
\end{equation}

\subsubsection{Time Evolution of mass eigenstates}

The time evolution of the mass eigenstates can be expressed in terms of the weak eigenstate for a produced initial weak eigenstate $D^0$ or $\Bar{D}^0$ as

\begin{equation}
    \begin{aligned}
        \ket{D^0(t)}&=f_+(t)\ket{D^0}+\frac{q}{p}f_-(t)\ket{\Bar{D}^0} \\
        \ket{\Bar{D}^0(t)}&=f_+(t)\ket{\Bar{D}^0}+\frac{p}{q}f_-(t)\ket{D^0},
    \end{aligned}
\end{equation}

where

\begin{equation}
\begin{aligned}
    f_{\pm}(t)=\frac{1}{2}e^{-iM_1t}&e^{-i\frac{1}{2}\Gamma_1t} \left(1\pm e^{-i\Delta Mt}e^{\frac{1}{2}\Delta \Gamma t}\right) \\
    \Delta M=M_2-&M_1,\hspace{1cm} \Delta \Gamma=\Gamma_1-\Gamma_2.
\end{aligned}
\end{equation}

The respective amplitudes $A(f)$ and $\Bar{A}(f)$ of the $D^0$ and $\Bar{D}^0$ decays into a final state $f$ are defined as

\begin{equation}
    A(f)=\mel{f}{H_{\Delta C=1}}{D^0}, \hspace{1cm} \Bar{A}(f)=\mel{f}{H_{\Delta C=1}}{\Bar{D}^0}.
\end{equation}

The amplitude ratios are useful to simplify notations, defined as

\begin{equation}
    \Bar{\rho}(f)=\frac{\Bar{A}(f)}{A(f)}=\frac{1}{\rho(f)}.
\end{equation}

As a consequence of the decay amplitudes, the decay times are obtained:

\begin{equation}
    \Gamma(D^0(t)\rightarrow f)\propto e^{-\Gamma_1t}\vert A(f)\vert^2\left(K_+(t)+K_-(t)\left\vert\frac{q}{p}\right\vert^2\vert\Bar{\rho}(f)\vert^2+2\Re{L^*(t)\left(\frac{q}{p}\right)\Bar{\rho}(f)}\right) 
    \label{eq:D0fdecayrate}
\end{equation}

\begin{equation}
    \Gamma(\Bar{D}^0(t)\rightarrow f)\propto e^{-\Gamma_1t}\vert \Bar{A}(f)\vert^2\left(K_+(t)+K_-(t)\left\vert\frac{p}{q}\right\vert^2\vert\rho(f)\vert^2+2\Re{L^*(t)\left(\frac{p}{q}\right)\rho(f)}\right),
    \label{eq:barD0fdecayrate}
\end{equation}

where

\begin{equation}
    \begin{aligned}
        \vert f_{\pm}(t)\vert^2&=\frac{1}{4}e^{-\Gamma_1t}K_{\pm}(t) \\
        f_-(t)f_+^*(t)&=\frac{1}{4}e^{-\Gamma_1t}L^*(t) \\
        K_{\pm}(t)&=1+e^{\Delta \Gamma t}\pm 2e^{\frac{1}{2}\Delta\Gamma t}\cos{\Delta Mt} \\
        L^*(t)&=1-e^{\Delta \Gamma t} + 2ie^{\frac{1}{2}\Delta\Gamma t}\sin{\Delta Mt}.
    \end{aligned}
    \label{eq:ffKL}
\end{equation}

Therefore, particle-antiparticle oscillations can contribute coherently for two different decay amplitudes, each decay with a relative weight that depends on the time of the decay (as a consequence of the oscillations)\bigi.

Hence, CP invariance is violated if the decay rate evolution for the decay of a neutral $D^0$ meson into a CP eigenstate $f_{\pm}$ is different from a single "pure" exponential\bigi. The mathematical equivalent of this statement is:

\begin{equation}
    \frac{d}{dt}e^{\Gamma t}\text{ rate}(D^0\rightarrow f_{\pm};t)\neq0\hspace{1cm}\forall \text{ real } \Gamma \implies\mathbf{CP}\text{ violation}.
\end{equation}

\subsection{CP violation mechanism}

\begin{figure}[ht]
    \vspace{-0.5cm}
    \centering
    \includegraphics[width=0.7\linewidth]{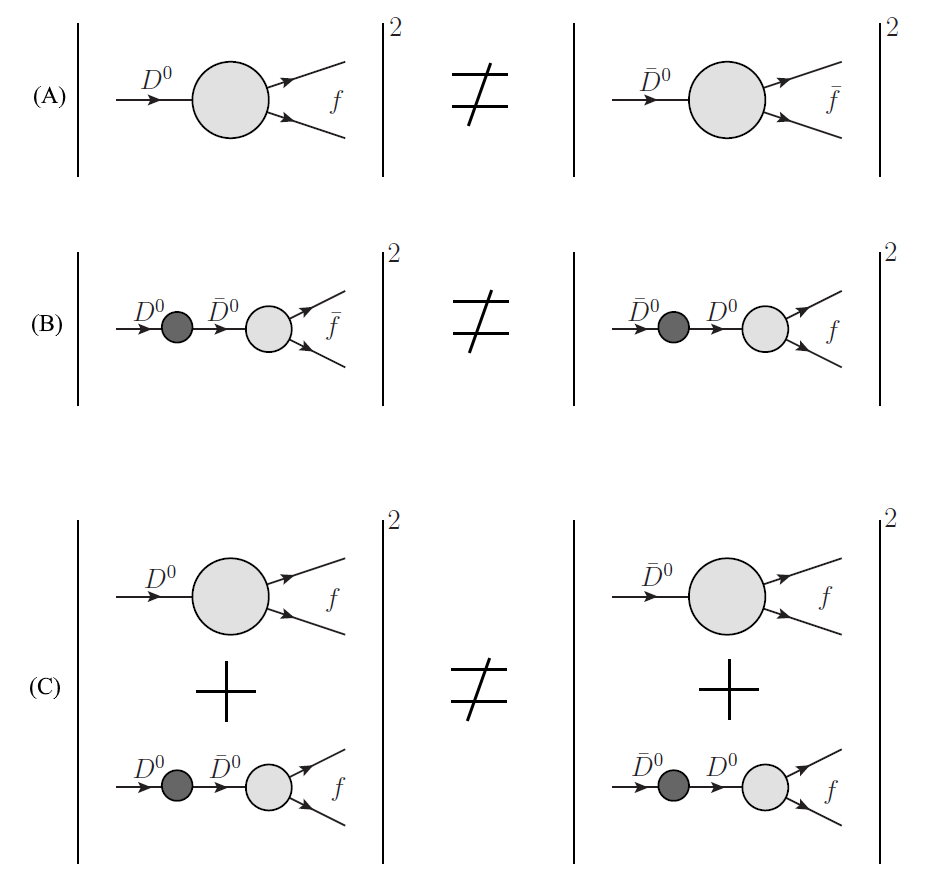}
    \caption{Diagram based on Figure 6.2 from Ref. \cite{bigi_sanda_2009}. The three CP violation mechanisms in quark decays from neutral mesons. (A) With no $D^0\leftrightarrow\Bar{D}^0$ mixing, there may be a CP asymmetry if the decay amplitudes $D^0\rightarrow f$ and $\Bar{D}^0\rightarrow \Bar{f}$ violate CP. (B) When there is mixing, CP violation can arise from flavour-specific final states. (C) If the final state $f$ is a CP eigenstate and both $D^0\rightarrow f$ and $\Bar{D}^0\rightarrow f$ occur, the two possible decay chains can interfere and generate a CP asymmetry.\bigi}
    \label{fig:classes}
\end{figure}

The decay rates are functions of various factors (including those which are consequences of oscillations), however, the occurrence of CP violation depends directly on it, since the difference in decay rates can be its cause. A physical explanation of this violation can be simplified by separating into simpler cases\bigi, visually represented in Figure \ref{fig:classes}.

\textbf{(A) Difference between $D^0\rightarrow f$ and $\Bar{D}^0\rightarrow f$ decay amplitudes and no $D^0\leftrightarrow \Bar{D}^0$ mixing}

Since no oscillations occur, the off-diagonal terms of the $\mathcal{H}$ matrix are zero:

\begin{equation}
    \Delta M=\Delta \Gamma =0.
\end{equation}

From equation \ref{eq:ffKL}, $K_+(t)\equiv 4$ and $K_-(t)\equiv L^*(t) \equiv 0$. Hence, equations \ref{eq:D0fdecayrate} and \ref{eq:barD0fdecayrate} show that the time evolutions of the decay rates are both dependent on $e^{-\Gamma_1t}$ and the condition for a CP asymmetry to arise is\bigi

\begin{equation}
    \vert A(f) \vert \neq \vert \Bar{A}(\Bar{f}) \vert
\end{equation}

and it can be measured directly as

\begin{equation}
    A_{CP}\equiv \frac{\Gamma(D^0(t)\rightarrow f)-\Gamma(\Bar{D}^0(t)\rightarrow \Bar{f})}{\Gamma(D^0(t)\rightarrow f)+\Gamma(\Bar{D}^0(t)\rightarrow \Bar{f})}=\frac{\vert A(f) \vert^2-\vert \Bar{A}(\Bar{f}) \vert^2}{\vert A(f) \vert^2+\vert \Bar{A}(\Bar{f}) \vert^2}.
\end{equation}

\textbf{(B) Flavour-specific decays into final states in the presence of $D^0\leftrightarrow \Bar{D}^0$ mixing}

Some flavour-specific final states can be a result of either a $D^0$ or a $\Bar{D}^0$ decay, but not from both, and are often provided by semileptonic decays\bigi:

\begin{equation}
    D^0\rightarrow l^++X \nleftarrow \Bar{D}^0, \text{  or  } D^0\nrightarrow l^-+X \leftarrow \Bar{D}^0
\end{equation}

In this situation, the impossibility of the decays are expressed by $A(l^-)=\Bar{A}(l^+)=0$ and X is a hadronic state. Equations \ref{eq:D0fdecayrate} and \ref{eq:barD0fdecayrate} yield:

\begin{equation}
    \begin{aligned}
        \Gamma(D^0(t)\rightarrow l^++X)&\propto e^{-\Gamma_1t}K_+(t)\vert A(l^+)\vert^2 \\
        \Gamma(D^0(t)\rightarrow l^-+X)&\propto e^{-\Gamma_1t}K_-(t)\left\vert \frac{q}{p}\right\vert^2\vert \Bar{A}(l^-)\vert^2 \\
        \Gamma(\Bar{D}^0(t)\rightarrow l^-+X)&\propto e^{-\Gamma_1t}K_+(t)\vert \Bar{A}(l^-)\vert^2 \\
        \Gamma(\Bar{D}^0(t)\rightarrow l^++X)&\propto e^{-\Gamma_1t}K_-(t)\left\vert \frac{p}{q}\right\vert^2\vert A(l^+)\vert^2
    \end{aligned}
    \label{eq:D0fb}
\end{equation}

If X corresponds to all hadronic states for this decay, and applying CPT invariance through $A(l^+)=\Bar{A}(l^-)$, the CP asymmetry in these semileptonic decays is independent of time, as measured in equation \ref{eq:ACPb}\bigi

\begin{equation}
    A_{SL}(D^0)\equiv \frac{\Gamma(D^0(t)\rightarrow l^-X)-\Gamma(\Bar{D}^0(t)\rightarrow l^+X)}{\Gamma(D^0(t)\rightarrow l^-X)+\Gamma(\Bar{D}^0(t)\rightarrow l^+X)}=\frac{1-\vert p/q\vert^4}{1+\vert p/q\vert^4}.
    \label{eq:ACPb}
\end{equation}

\textbf{(C) Flavour-non-specific decays with different rates}

These decays are allowed for both $D^0$ and $\Bar{D}^0$ (equation \ref{eq:D0fBarD0}), resulting in two possible decay chains due to $D^0\leftrightarrow\Bar{D}^0$ mixing. The interference between these two chains can result in the CP asymmetry. This case must be separated in two cases, one in which $\vert A(f)\vert=\vert\Bar{A}(f)\vert$ and the other when $\vert A(f)\vert\neq\vert\Bar{A}(f)\vert$

\begin{equation}
    D^0\rightarrow f \leftarrow \Bar{D}^0
    \label{eq:D0fBarD0}
\end{equation}

Some of these possible decays for the $D^0$ meson are $D^0\rightarrow K\Bar{K}, \pi\Bar{\pi}, K\pi, \Bar{K}\pi\leftarrow\Bar{D}^0$, of which only the former two are CP eigenstates.

\textbf{(C1)} $\vert A(f)\vert=\vert\Bar{A}(f)\vert$

Even if the amplitudes are equal, it is possible that $\vert q/p\vert=1$ and $\vert\Bar{\rho}(f)\vert=1$, conditions that, when substituted in equations \ref{eq:D0fdecayrate} and \ref{eq:barD0fdecayrate} result in the following decay rates 

\begin{equation}
    \begin{aligned}
        \Gamma(D^0(t)&\rightarrow f)\propto 2e^{-\Gamma_1t}\vert A(f)\vert^2 \\
        &\times \left(1+e^{\Delta \Gamma t}+\Re{\frac{q}{p}\Bar{\rho}(f)}(1-e^{\Delta \Gamma t}) - 2\Im{\frac{q}{p}\Bar{\rho}(f)}e^{\frac{1}{2}\Delta\Gamma t}\sin{\Delta Mt}\right) 
    \end{aligned}
    \label{eq:D0fc1}
\end{equation}

\begin{equation}
    \begin{aligned}
        \Gamma(\Bar{D}^0(t)&\rightarrow f)\propto 2e^{-\Gamma_1t}\vert A(f)\vert^2 \\
        &\times \left(1+e^{\Delta \Gamma t}+\Re{\frac{q}{p}\Bar{\rho}(f)}(1-e^{\Delta \Gamma t}) + 2\Im{\frac{q}{p}\Bar{\rho}(f)}e^{\frac{1}{2}\Delta\Gamma t}\sin{\Delta Mt}\right) 
    \end{aligned}
    \label{eq:BarD0fc1}
\end{equation}

from which the CP asymmetry can be measured (equation \ref{eq:ACPc1}) by using the fact $\frac{q}{p}\Bar{\rho}(f)=(\frac{p}{q}\rho(f))^*$, and rewriting $q/p=e^{i\phi_{\Delta C=2}}$ and $\Bar{\rho}=e^{i\phi_{\Delta C=1}}$

\begin{equation}
    A_{CP}\equiv \frac{\Gamma(D^0(t)\rightarrow f)-\Gamma(\Bar{D}^0(t)\rightarrow f)}{\Gamma(D^0(t)\rightarrow f)+\Gamma(\Bar{D}^0(t)\rightarrow f)}=\frac{-2\sin(\phi_{\Delta C=2}+\phi_{\Delta C=1})e^{\frac{1}{2}\Delta \Gamma t}\sin(\Delta Mt)}{1+e^{\Delta \Gamma t}+\cos(\phi_{\Delta C=2}+\phi_{\Delta C=1})(1-e^{\Delta \Gamma t})}.
    \label{eq:ACPc1}
\end{equation}

Under this condition, a CP asymmetry may arise if $A_{CP}\neq0$, which happens when $D^0\leftrightarrow\Bar{D}^0$ oscillations occur (generating $\Delta M\neq0$), as well as $\phi_{\Delta C=2}+\phi_{\Delta C=1}\neq0$, equivalent to the sum of the arguments of $q/p$ and $\Bar{\rho}(f)$ differing from zero.

\textbf{(C2)} $\vert A(f)\vert\neq\vert\Bar{A}(f)\vert$

In this situation, the following decay rates are obtained 

\begin{equation}
    \begin{aligned}
        &\Gamma(D^0(t)\rightarrow f)\propto \frac{1}{2}e^{-\Gamma_1t} G_f(t) \\
        G_f(t) =\vert A(f)\vert^2 &\left(1+ \vert \Bar{\rho}(f)\vert^2 + (1- \vert \Bar{\rho}(f)\vert^2)\cos{\Delta Mt} - 2\Im{\frac{q}{p}\Bar{\rho}(f)}\sin{\Delta Mt}\right) 
    \end{aligned}
\end{equation}

\begin{equation}
    \begin{aligned}
        &\Gamma(\Bar{D}^0(t)\rightarrow f)\propto \frac{1}{2}e^{-\Gamma_1t}\vert \Bar{G}_f(t)\\
        \Bar{G}_f(t)=\Bar{A}(f)\vert^2 &\left(1+ \vert \rho(f)\vert^2 + (1- \vert \rho(f)\vert^2)\cos{\Delta Mt} - 2\Im{\frac{p}{q}\rho(f)}\sin{\Delta Mt}\right). 
    \end{aligned}
\end{equation}

Consequently, the CP asymmetry is

\begin{equation}
    \begin{aligned}
        \frac{G_f(t)-\Bar{G}_f(t)}{G_f(t)+\Bar{G}_f(t)}=&C_f\cos{\Delta M t}-S_f\sin{\Delta M t} \\
        C_f=\frac{1-\vert \Bar{\rho}(f)\vert^2}{1+\vert \Bar{\rho}(f)\vert^2},& \hspace{1cm} S_f=\frac{2\Im{\frac{q}{p}\Bar{\rho}(f)}}{1+\vert \Bar{\rho}(f)\vert^2}.
    \end{aligned}
\end{equation}

This asymmetry has two clear separate sources, given by the coefficients $C_f$ and $S_f$, constrained only by $C_f^2+S_f^2\le1$. The asymmetry from the coefficient $C_f$ has a more evident physical interpretation, resulting from $\vert \Bar{\rho}(f)\vert^2\neq1$, the situation in which the decay amplitudes are different, similar to the first described case of direct CP violation.

\subsection{Time-integrated observables}

The different cases of CP asymmetries generated by varying decay times may not always be directly observable, since the decay time of short-lived particles can be extremely difficult to access experimentally. Typical experiments often use time-integrated versions of the previously discussed predictions: detection is made by integrating over all times of decay\bigi.

Defining the oscillation rates in comparison to the decay rates as

\begin{equation}
    x=\frac{\Delta M}{\Gamma}, \hspace{1cm} y=\frac{\Delta \Gamma}{2\Gamma}, \hspace{1cm} \Gamma=\frac{1}{2}(\Gamma_1+\Gamma_2),
\end{equation}

the general decay rates from equations \ref{eq:D0fdecayrate} and \ref{eq:barD0fdecayrate} can be expressed in terms of

\begin{equation}
    \begin{aligned}
        &\expval{K_{\pm}}\equiv\int_0^{\infty} e^{-\Gamma_1t}K_{\pm}(t)dt = \frac{2}{\Gamma}\left(\frac{1}{1-y^2}\pm \frac{1}{1+x^2}\right) \\
        & \expval{L^*}\equiv\int_0^{\infty} e^{-\Gamma_1t}L^*(t)dt = \frac{2}{\Gamma}\left(-\frac{y}{1-y^2}+ i\frac{x}{1+x^2}\right).
    \end{aligned}
\end{equation}

Time-integrated versions of the expressions in equation \ref{eq:D0fb} are

\begin{equation}
    \begin{aligned}
        \int_0^{\infty}\Gamma(D^0(t)\rightarrow l^++X)dt&\propto \frac{2+x^2-y^2}{(1-y^2)(1+x^2)}\vert A(l^+)\vert^2 \\
        \int_0^{\infty}\Gamma(D^0(t)\rightarrow l^-+X)dt&\propto \frac{x^2+y^2}{(1-y^2)(1+x^2)}\left\vert \frac{q}{p}\right\vert^2\vert \Bar{A}(l^-)\vert^2 \\
        \int_0^{\infty}\Gamma(\Bar{D}^0(t)\rightarrow l^-+X)dt&\propto \frac{2+x^2-y^2}{(1-y^2)(1+x^2)}\vert \Bar{A}(l^-)\vert^2 \\
        \int_0^{\infty}\Gamma(\Bar{D}^0(t)\rightarrow l^++X)dt&\propto \frac{x^2+y^2}{(1-y^2)(1+x^2)}\left\vert \frac{p}{q}\right\vert^2\vert A(l^+)\vert^2.
    \end{aligned}
\end{equation}

Equations \ref{eq:D0fc1} and \ref{eq:BarD0fc1} can also be rewritten for $y=\frac{\Delta \Gamma}{2\Gamma}=0$, maintaining the constraint $\vert\Bar{\rho}\vert=\left\vert\frac{q}{p}\right\vert=1$:

\begin{equation}
    \int_0^{\infty}\Gamma(D^0(t)\rightarrow f)dt\propto 4\vert A(f)\vert^2 \left(1-\frac{x}{1+x^2}\Im{\frac{q}{p}\Bar{\rho}(f)}\right) 
\end{equation}

\begin{equation}
    \int_0^{\infty}\Gamma(\Bar{D}^0(t)\rightarrow f)dt\propto 4\vert A(f)\vert^2 \left(1+\frac{x}{1+x^2}\Im{\frac{q}{p}\Bar{\rho}(f)}\right).
\end{equation}

Hence, in the \textbf{(C1)} case, since the asymmetry is due to the interference between decay rates and oscillation amplitudes, the ratio $x=\frac{\Delta M}{\Gamma}$ between the two is the factor on which the CP asymmetry depends. 

\section{Experimental Discovery}
\label{sec:expdis}

The experimental confirmation of $\mathbf{CP}$ violation in charm decays with $5\sigma$ confidence level was in 2019 at the \textit{LHCb} experiment, as reported by Ref. \cite{observation_cpv}. The SM prediction for the size of such violation is of the order $10^{-3}-10^{-4}$\article, but there could be some Beyond the Standard Model (BSM) predicted particles capable of altering this size. As described in the article, the measurements made regard the $D^0\rightarrow K^+K^-$ and $D^0\rightarrow\pi^+\pi^-$ decays, evidently made in comparison to the $\Bar{D}^0\rightarrow K^+K^-$ and $\Bar{D}^0\rightarrow\pi^+\pi^-$ decays.

The observation was performed through the difference in time-integrated decays by using proton-proton collision data collected by the LHCb detector at a center-of-mass energy of $13$ TeV correspondent to an integrated luminosity of $5.9$ $fb^{-1}$\article. The final states of the decays are CP eigenstates with zero charm ($C=0$), produced from $D^0$ and $\Bar{D}^0$ decays. The time-dependent CP asymmetry between the final states $f$ at time $t$ can be measured through equation \ref{eq:ACP} 

\begin{equation}
    A_{CP}(f; t)\equiv\frac{\Gamma(D^0(t)\rightarrow f)-\Gamma(\Bar{D}^0(t)\rightarrow f)}{\Gamma(D^0(t)\rightarrow f)+\Gamma(\Bar{D}^0(t)\rightarrow f)}.
    \label{eq:ACP}
\end{equation}

For the final states being tested in decays of a neutral particle such as $D^0$, the asymmetry falls under the discussed category \textbf{(C)}, generated by the interference between a direct CP violation in the decay amplitudes and the $D^0\leftrightarrow\Bar{D}^0$ oscillations. As mentioned, \textit{LHCb} was capable of determining the time-integrated CP asymmetry $A_{CP}(f)$, which can be written as:

\begin{equation}
    A_{CP}(f)\approx a_{CP}^{dir}(f)-\frac{\expval{t(f)}}{\tau(D^0)}A_{\Gamma}(f).
    \label{eq:expACP}
\end{equation}

Through the mean decay time of $D^0\rightarrow f$ decays in the reconstructed sample ($\expval{t(f)}$), this expression takes in consideration the dependence on the variation of the reconstruction efficiency as a function of the decay time\article, thus incorporating the effects of the time-dependent experimental efficiency\article. The $a_{CP}^{dir}(f)$ term corresponds to the direct CP asymmetry, $\tau(D^0)$ the lifetime of $D^0$ and $A_{\Gamma}(f)$ the asymmetry in the effective decay widths of $D^0\rightarrow f$ and $\Bar{D}^0\rightarrow f$.

The usage of both $\pi^+\pi^-$ and $K^+K^-$ decay channels is an experimental tool used to maximize the observed CP asymmetry. As observed in the diagrams from Figures \ref{fig:DpiT}, \ref{fig:DpiP}, \ref{fig:DKT} and \ref{fig:DKP}, the interaction vertex for the pion decays involve a $c$ quark going to a $d$ quark, while for the kaon decays, the $c$ quark goes to an $s$ quark. However, U-spin-symmetry between the strange and down quarks dictate the charm decays into either of them yield direct CP asymmetries equal in magnitude but opposite in sign. Hence, by taking the difference between these asymmetries, the total observed CP asymmetry is increased. Taking $A_{\Gamma}$ to be independent of the final state, 
the difference in the asymmetries is expressed as\article:

\begin{equation}
    \Delta A_{CP} \equiv A_{CP}(K^-K^+)-A_{CP}(\pi^-\pi^+)\approx \Delta a_{CP}^{dir}-\frac{\Delta\expval{t}}{\tau(D^0)}A_{\Gamma},
\end{equation}

where $\Delta a_{CP}^{dir}\equiv a_{CP}^{dir}(K^-K^+)-a_{CP}^{dir}(\pi^-\pi^+)$ and $\Delta\expval{t}=\expval{t(K^-K^+)}-\expval{t(\pi^-\pi^+)}$.

The production of the $D^0$ or $\Bar{D}^0$ eigenstates are achieved at \textit{LHCb} through flavour tagging with muons and pions. These eigenstates are produced either at a proton-proton collision point in the strong decay of a $D^{*+}$ meson into a $D^{0}\pi^+$ pair, or at a vertex from the semileptonic decay of a hadron ($\Bar{B}$) containing a $b$ quark given by $\Bar{B}\rightarrow D^0\mu^-\Bar{\nu}_{\mu}X$. The charge of the pion in the $D^{*+}$ decay determines the flavor of the produced $D^0$, thus being $\pi$-tagged. The charge of the muon in the $\Bar{B}$ hadron determines the flavor of the $D^0$ from this $\mu$-tagged decay\article. Therefore, naming $N$ the measured signal for a given decay, the asymmetries in these channels are

\begin{equation}
    \begin{aligned}
        A_{raw}^{\pi-tagged}(f)&\equiv \frac{N(D^{*+}\rightarrow D^0(f)\pi^+)-N(D^{*-}\rightarrow \Bar{D}^0(f)\pi^-)}{N(D^{*+}\rightarrow D^0(f)\pi^+)+N(D^{*-}\rightarrow \Bar{D}^0(f)\pi^-)} \\
        A_{raw}^{\mu-tagged}(f)&\equiv \frac{N(\Bar{B}\rightarrow D^0(f)\mu^-\Bar{\nu}_{\mu}X)-N(B\rightarrow \Bar{D}^0(f)\mu^+\nu_{\mu}X)}{N(\Bar{B}\rightarrow D^0(f)\mu^-\Bar{\nu}_{\mu}X)+N(B\rightarrow \Bar{D}^0(f)\mu^+\nu_{\mu}X)}.
    \end{aligned}
\end{equation}

The $\pi$-tagged($\mu$-tagged) signal generated approximately $44$ $(9)$ million $D^0\rightarrow K^+K^-$ decays and $14$ $(3)$ million $D^0\rightarrow \pi^+\pi^-$ decays\article. These numbers contain both $D^0$ and $\Bar{D}^0$ decays.

The uncertainties of the experimental setup are taken in consideration within the statistical uncertainty while uncertainties associated to contamination from other decays are accounted for within systematic uncertainties. The measured results and calculated uncertainties are obtained for both $\pi$-tagged and $\mu$-tagged signals as\article

\begin{equation}
    \begin{aligned}
        \Delta A_{CP}^{\pi-tagged}=&[-18.2\pm3.2(stat.)\pm0.9(syst)]\cdot10^{-4} \\
        \Delta A_{CP}^{\mu-tagged}=&[-9\pm8(stat.)\pm5(syst)]\cdot10^{-4}.
    \end{aligned}
\end{equation}

Such results were combined with previous \textit{LHCb} measurements, yielding 

\begin{equation}
    \Delta A_{CP}=[-15.4\pm2.9]\cdot10^{-4},
\end{equation}

with a $5.3\sigma$ deviation from zero\article.

Current knowledge of the mean $D^0$ lifetime and of the $D^0\rightarrow K^-K^+$ and $D^0\rightarrow \pi^-\pi^+$ mean decay times result in $\Delta\expval{t}/\tau (D^0)=0.115\pm0.002$. The \textit{LHCb} experiment has also measured the average $A_{\Gamma}=(-2.8\pm2.8)\cdot10^{-4}$. Hence, from equation \ref{eq:expACP}, the direct CP violation term can be calculated as $\Delta a_{CP}^{dir}=(-15.7\pm2.9)\cdot10^{-4}$. Such result allows the interpretation that $\Delta A_{CP}$ in the studied decay channels are mainly due to the contribution of direct CP violation. 

An important fact to be pointed out is that the SM predicts the interference between penguin and tree diagrams to be the source of CP violation, but given the estimated ratio between penguin and tree diagrams $\vert P/T\vert\approx0.1$ with an approach from QCD, the upper bound of the CP asymmetry is $\Delta A_{CP}\lesssim2.6\times 10^{-4}$\textsuperscript{\cite{review}}, around one order of magnitude smaller than the observed result.

\section{CP violation in neutrino oscillations}

Similarly to its violation in quark decays and neutral meson decays, CP can be violated in neutrino oscillations, as described by the PMNS matrix, equivalent to the CKM matrix in various ways. Neutrinos are described in the Standard Model as ultra-relativistic particles that interact exclusively through the weak force. Analogously to quarks, neutrinos have mass eigenstates differing from the weak eigenstates, and when propagating in space as their mass eigenstates, neutrinos may be detected at a different position with a flavor different from its original flavor. 

This phenomenon is explained through the PMNS matrix, which correlates the neutrino mass and weak eigenstates:

\begin{equation}
    \begin{pmatrix}
        \nu_e \\ \nu_{\mu} \\ \nu_{\tau} 
    \end{pmatrix}
    =
    \begin{pmatrix}
        U_{e1} & U_{e2} & U_{e3}\\ U_{\mu1} & U_{\mu2} & U_{\mu3}\\ U_{\tau1} & U_{\tau2} & U_{\tau3}
    \end{pmatrix}
    \begin{pmatrix}
        \nu_1 \\ \nu_{2} \\ \nu_{3} 
    \end{pmatrix}.
\end{equation}

This matrix possesses the same unitarity properties of the CKM matrix, allowing for the same parametrization using three mixing angles and one CP violating complex phase

\begin{equation}
    U_{PMNS}=
    \begin{pmatrix}
    U_{e1} & U_{e2} & U_{e3} \\
    U_{\mu1} & U_{\mu2} & U_{\mu3} \\
    U_{\tau1} & U_{\tau2} & U_{\tau3} 
    \end{pmatrix}
    =
    \begin{pmatrix}
    c_{12}c_{13} & s_{12}c_{13} & s_{13}e^{-i\delta} \\
    -s_{12}c_{23}-c_{12}s_{23}s_{13}e^{i\delta} & c_{12}c_{23}-s_{12}s_{23}s_{13}e^{i\delta} & s_{23}c_{13} \\
    s_{12}s_{23}-c_{12}c_{23}s_{13}e^{i\delta} & -c_{12}s_{23}-s_{12}c_{23}s_{13}e^{i\delta} & c_{23}c_{13} 
    \end{pmatrix}.
    \label{eq:PMNS}
\end{equation}

The phenomenon of CP violation is neutrino oscillations can be seen from the oscillation probabilities for neutrinos. A neutrino originated as a weak eigenstate can be represented as a linear combination of the mass eigenstates: 

\begin{equation}
\begin{aligned}
    \label{eq:lin}
    \ket{\nu_{\alpha}}&=\sum_{k} U_{\alpha k}^*\ket{\nu_k} \hspace{2cm}      (\alpha = e, \mu, \tau).
\end{aligned}
\end{equation}

It propagates through space as its mass eigenstate, obeying Schrödinger's equation (equation \ref{eq:schro}). However, ultra-relativistic neutrinos may be approximated as plane waves, characterized by the time-evolution\giunti:

\begin{equation}
    \ket{\nu_k (t)}=e^{-i E_k t}\ket{\nu_k}.
    \label{eq:evt}
\end{equation}

Hence, the time-evolved weak eigenstate is

\begin{equation}
    \ket{\nu_{\alpha}(t)}=\sum_{k} U_{\alpha k}^*e^{-i E_k t}\ket{\nu_k}.
\end{equation}

Rewriting the mass eigenstates as a linear combination of the weak eigenstates with the conjugate of the PMNS matrix coefficients

\begin{equation}
    \ket{\nu_k}=\sum_{\alpha} U_{\alpha k}\ket{\nu_{\alpha}}.
    \label{eq:lininv}
\end{equation}

The dispersion relation for ultra-relativistic neutrinos, allows the approximations $E_k-E_j \simeq \frac{\Delta m_{kj}^2}{2E}$ (where $\Delta m_{kj}^2\equiv m_{k}^2 - m_{j}^2$)\giunti and $t\approx L$. Thus, the oscillation probability as a function of neutrino energy and distance traveled is

\begin{equation}
    P_{\nu_{\alpha}\rightarrow \nu_{\beta}}(L, E)=|\braket{\nu_{\beta}}{\nu_{\alpha}(t)}|^2=\sum_{k, j}U_{\alpha k}^* U_{\beta k} U_{\alpha j} U_{\beta j}^* exp\left(-i\frac{\Delta m_{kj}^2 L}{2E}\right).
\end{equation}

The same deduction can be made for antineutrinos, which enter the interaction vertex as a spinor, rather than the adjoint spinor, hence respecting the relations in equation \ref{eq:antineut} and having the oscillation probability in equation \ref{eq:probanti}

\begin{equation}
    \ket{\bar{\nu}_{\alpha}}=\sum_{k} U_{\alpha k}\ket{\bar{\nu}_k} \hspace{1cm} \text{and} \hspace{1cm} \ket{\bar{\nu}_{k}}=\sum_{k} U_{\alpha k}^*\ket{\bar{\nu}_{\alpha}} \hspace{2cm} (\alpha = e, \mu, \tau)
    \label{eq:antineut}
\end{equation}

\begin{equation}
    P_{\Bar{\nu}_{\alpha}\rightarrow \Bar{\nu}_{\beta}}(L, E)=\sum_{k, j}U_{\alpha k} U_{\beta k}^* U_{\alpha j}^* U_{\beta j} exp\left(-i\frac{\Delta m_{kj}^2 L}{2E}\right).
    \label{eq:probanti}
\end{equation}

Since the coefficients multiplying the exponent within the summation are now conjugate in relation to the probability for neutrinos, a non-zero complex phase in the PMNS matrix yields different oscillation probabilities for neutrino and antineutrino oscillations. This difference can be more explicitly seen when the probabilities are rewritten as in equations \ref{eq:neutprobs} and \ref{eq:aneutprobs}.

\begin{equation}
    \begin{aligned}
        P_{\nu_{\alpha}\rightarrow \nu_{\beta}}(L, E)=\delta_{\alpha \beta} -4\sum_{k> j}\Re{U_{\alpha k}^* U_{\beta k} U_{\alpha j} U_{\beta j}^*} \sin^2\left(\frac{\Delta m_{kj}^2 L}{4E}\right)\\ +2\sum_{k> j}\Im{U_{\alpha k}^* U_{\beta k} U_{\alpha j} U_{\beta j}^*} \sin \left(\frac{\Delta m_{kj}^2 L}{2E}\right), 
        \label{eq:neutprobs}
    \end{aligned}
\end{equation}

\begin{equation}
    \begin{aligned}
        P_{\Bar{\nu}_{\alpha}\rightarrow \Bar{\nu}_{\beta}}(L, E)=\delta_{\alpha \beta} -4\sum_{k> j}\Re{U_{\alpha k}^* U_{\beta k} U_{\alpha j} U_{\beta j}^*} \sin^2\left(\frac{\Delta m_{kj}^2 L}{4E}\right)\\ -2\sum_{k> j}\Im{U_{\alpha k}^* U_{\beta k} U_{\alpha j} U_{\beta j}^*} \sin \left(\frac{\Delta m_{kj}^2 L}{2E}\right).
        \label{eq:aneutprobs}
    \end{aligned}
\end{equation}

Henceforth, the $\delta_{CP}$ factor defines the violation of CP symmetry, given that, if $\delta_{CP}\neq 0$, then $P_{\nu_{\alpha}\rightarrow \nu_{\beta}} \neq P_{\bar{\nu}_{\alpha}\rightarrow \bar{\nu}_{\beta}}$, yielding a CP asymmetry of

\begin{equation}
    A_{CP}^{\alpha \beta}=P_{\nu_{\alpha}\rightarrow \nu_{\beta}}-P_{\Bar{\nu}_{\alpha}\rightarrow \Bar{\nu}_{\beta}}.
\end{equation}

\section{Discussion}

The phenomenon of CP violation is evidently present in the dynamics of several elementary particles interacting via the weak force. This violation was defined by Sakharov as one of the conditions for baryogenesis, however it is not sufficient to explain the experimental observations. In all cases discussed, the weak force, via an interference with the strong force, is the source of the CP violation in $D^0$ decays. 

The violation was also observed to be rather small when compared to other systems, as justified by the order of magnitude of the complex phase in the CKM elements, seen more clearly via the Wolfenstein parametrization. In spite of the size of the CP asymmetries, it has more recently been probed experimentally in the $D^0$ system, much later than in kaons and $B$-meson systems. Up to date, neutrino oscillation experiments show evidences of CP violation, but the off-diagonal elements of the PMNS matrix indicate a much greater mixing and possibly a larger CP asymmetry. 

The analysis of the D-meson system that was carried out demonstrates the effects of the complex phase in the CKM matrix in generating direct CP violation as well as the interference of neutral meson mixing contributing to this violation. 

Even though different contexts are mentioned, the framework for their description is extremely similar. The difference between mass and weak eigenstates is the fundamental pillar on which this framework is based: the description of the linear relations between eigenstates through non-diagonal matrices only have the physical requirement of unitarity, but does not constrain the nature of the particle fields. The existence of three fermion generations, both for leptons and quarks imply the existence of the fundamental irreducible complex phase in the CKM and PMNS matrices. Since baryonic antimatter enters any interaction vertex as the adjoint spinor, the CP conjugate processes are described by the complex conjugates of these phases, thus allowing for the interference between the inverted weak phase and the unchanged strong phase shift in the meson system.

As described, the constraints on $D^0\leftrightarrow \Bar{D}^0$ oscillations, along with constraints on decay amplitudes can combine and create different CP violation mechanisms, conditioned by $\vert\frac{q}{p}\vert$ and $\rho(f)$.

The experimental confirmation of CP violation in charm decays is not in accordance with SM predictions, which indicate an upper bound\textsuperscript{\cite{review}} for the CP asymmetry of $\Delta A_{CP}\lesssim2.6\times 10^{-4}$. A possible explanation regards a dynamical enhancement of the penguin diagram amplitudes in comparison to tree diagram amplitudes, with predictions attempting to include non perturbative effects\textsuperscript{\cite{review}}. These non perturbative effects could contribute to a larger strong phase interfering with the weak phase, hence increasing the upper bound for the CP asymmetry, so as to reach agreement with the experimental observations.

A different possible explanation involves final state interactions of hadronic order. These strong final state interactions could be the cause for a direct enhancement of the CP asymmetry, rather than the enhancement of the penguin contribution.

\section{Conclusion}

Independently of the combination of phenomena violating CP or whether this description is for quark flavour changing processes and meson oscillations, or neutrino oscillations, the Standard Model description of a complex phase in a unitary matrix relating free-particle and interaction eigenstates is an essential framework, acting as a source for CP violation when interfering with a strong phase in the presence of two possible decay paths.

However, the expected results for this description diverge from the experimental observations. Within the current framework, an enhancement of penguin diagrams might explain this difference, or perhaps final state interactions of hadronic order could motivate an enhancement in the predicted CP asymmetry. There are also proponents of New Physics motivating the discrepancies.

Regardless of these possibilities, further measurements of CP violation in the D-meson system will help to figure out if the size of the CP asymmetry is compatible or not with its explanations within the Standard Model.

\newpage

\bibliographystyle{unsrt}
\bibliography{bibliography}

\end{document}